\documentclass[rmp,aps,twocolumn,nofootinbib,endfloats]{revtex4}

\usepackage{amsmath}

\usepackage{amsfonts}

\usepackage{graphicx}

\usepackage{graphics}
\usepackage{amsmath}

\usepackage{epsfig}

\renewcommand{\vec}[1]{\mbox{\boldmath$#1$}}

\newcommand{\dif}{\mathrm{d}}

\newcounter{saveeqn}

\begin{document}
\title{Scaled-Particle Theory and the Length-scales Involved in
Hydrophobic Hydration of Aqueous Biomolecular Assemblies}

\author{Henry S. Ashbaugh}
\affiliation{Theoretical Division, Los Alamos National Laboratory, 
Los Alamos, NM 87545, USA}
\author{Lawrence  R.  Pratt}
\affiliation{Theoretical Division, Los Alamos National Laboratory, 
Los Alamos, NM 87545, USA}

\preprint{LA-UR-03-2144}

\begin{abstract}  
Hydrophobic hydration plays a crucial role in self-assembly processes
over multiple length-scales, from the microscopic origins of inert gas
solubility in water, to the  mesoscopic organization of proteins and
surfactant structures, to macroscopic phase separation. Many theoretical
studies focus on the molecularly detailed interactions between oil and 
water, but the extrapolation of molecular-scale models to larger
length-scale hydration phenomena is sometimes not  warranted.
Scaled-particle theories are based upon an interpolative view of that
microscopic$\rightarrow$macroscopic issue.  We revisit the
scaled-particle theory proposed thirty years ago by Stillinger (\emph{J.
Soln. Chem.} {\bf 2}, 141-158 (1973)), adopt a practical generalization,
and consider the implications for hydrophobic hydration in light of our
current understanding. The generalization is based upon identifying a
molecular length, implicit in previous applications of scaled-particle
models, that provides an effective radius for  joining microscopic and
macroscopic descriptions.  We demonstrate that the generalized theory
correctly reproduces many of the anomalous thermodynamic properties of
hydrophobic hydration for molecularly sized solutes, including
solubility minima and entropy convergence, successfully interpolates
between the  microscopic and macroscopic extremes, and provides new
insights into the underlying molecular mechanisms. The model considered
here serves as a reference for theories that bridge microscopic and
macroscopic hydrophobic effects.  The results are discussed in terms of
length-scales associated with component phenomena; in particular we
first discuss the  micro-macroscopic joining radius identified by the
theory, then we discuss in turn the Tolman length that leads to an
analogous length describing curvature corrections of a surface area
model of hydrophobic hydration free energies, and the length-scales on
which \emph{entropy convergence} of hydration free  energies are
expected.
\end{abstract}                                                                 

\date{\today}
\maketitle
\tableofcontents

\section{Introduction}\label{intro}

The  adage ``oil and water don't mix'' dominates thinking about
hydrophobic effects that are upheld, nearly universally, as the primary
thermodynamic impetus for a number of important aqueous solution
phenomena, including the environmental fate of pollutants, surfactant
assembly, biological membrane formation, and the folding of globular
proteins \cite{Kauzmann:59,Tanford:80,Blokzijl:AC:93,SimonsonT:Eleadp}.
Enigmatic temperature signatures  --- such as the fact that many soluble
proteins unfold both upon heating {\em and} cooling --- offer the
primary puzzles of hydrophobic effects, and are characteristic of the
aqueous milieu.   The ability to reproduce these temperature signatures
from basic principles is essential for understanding the temperature
range of functional behavior of biophysical structures, and of aqueous
phase nanotechnology designed by analogy with the molecular machinery of
biophysics.  Important aspects of these puzzles are that the hydrophobic
temperature signatures are strongly affected by the spatial
length-scales of the hydrophobic solution structures.  This review
focuses  on recent progress in unraveling the puzzles of temperature
signatures and length-scales characteristic of hydrophobic hydration.

\vfil

\subsection{An empirical length characteristic of a liquid in
coexistence with a dilute vapor}\label{EWsubsection} 
\begin{figure}[h]
\includegraphics[scale=0.45]{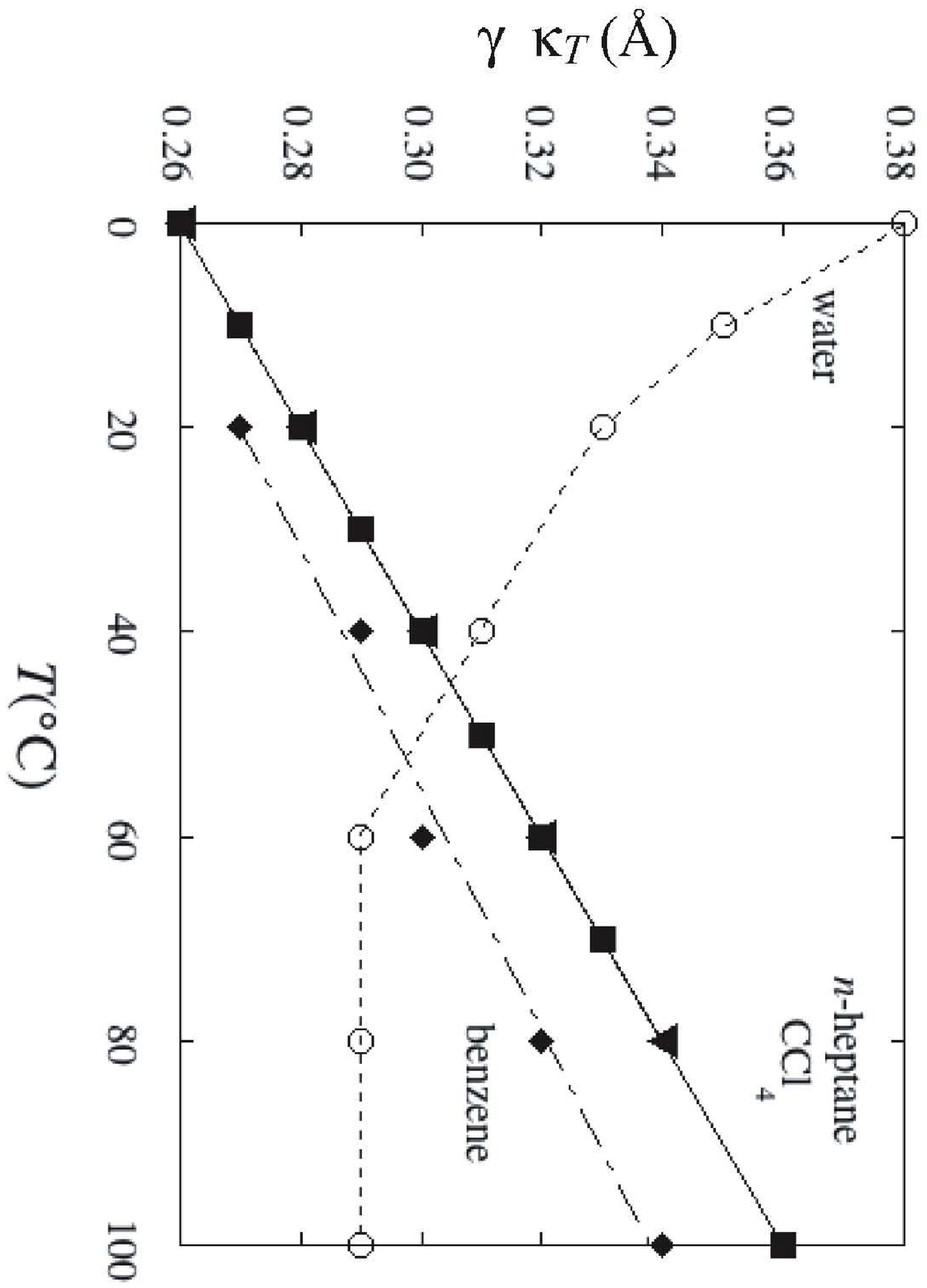}
\caption{Product of the liquid-vapor surface tension, $\gamma$, and the
the isothermal compressibility, $\kappa_T\equiv -
\frac{1}{V}\left(\frac{\partial V}{\partial p}\right)_T$, for several
liquid solvents at low temperatures so that the density of the
coexisting vapor is low.  Even though the individual factors differ
substantially in magnitude, this product is a length characteristic of
the liquid, accessible on the basis of macroscopic measurements, and can
be taken as proportional to a molecular correlation length.  The
interesting observation here is that temperature dependence of this
correlation length is qualitatively different for liquid water than for
these organic solvents.}
\label{figEW}
\end{figure}

The calling-out of
a particular length-scale, and the role it plays in statistical theories
of solutions, has been a primary feature of recent discussions of
hydrophobic effects \cite{Lee:85,Lum:99}, though the length-scales noted
in those two cases were different from each other. It is interesting,
therefore, to consider length-scale issues more broadly.    We can start
by noting the classic suggestion of \cite{EW70} for a length-scale
characteristic of a liquid in coexistence with a dilute vapor phase,
namely the product of the liquid-vapor interfacial tension, $\gamma$,
and the isothermal compressibility of the liquid, $\kappa_T\equiv -
\frac{1}{V}\left(\frac{\partial V}{\partial p}\right)_T$;  see
Fig.~\ref{figEW}.  The argument supporting this suggestion was physical,
heuristic, and our discussion below of the scaled-particle theories will
shed some additional light on this length-scale. It was immediately
observed, however, that away from a critical region $\gamma  \kappa_T$
exhibited limited variation from liquid to liquid though the individual
factors could differ by more than two orders of magnitude. 

Fig.~\ref{figEW} shows that the temperature dependence of the length
$\gamma \kappa_T$ is qualitatively different for liquid water than it is
for the organic solvents shown. This qualitatively different behavior is
mostly ascribable to the fact that the compressibility of liquid water
displays a minimum at 46$^\circ$C; it decreases with increasing
temperature for temperatures lower than this, and has a smaller net
variation over this temperature domain compared to the other solvents.
Though liquid water is much less compressible than are the other
liquids, the liquid-vapor surface tension is much higher for water than 
for the other cases. This exemplifies the point that the product $\gamma
\kappa_T$ has smaller variations than do the individual factors. Still,
liquid water is distinguished from liquids generically by peculiar
temperature dependences. Similarly, the peculiarities of 
hydrophobic effects are  temperature dependences that imply
entropic stabilization of conformations and assemblies of hydrophobic
solutes in aqueous solutions.

The theoretical description of hydrophobic effects has recently
progressed markedly, and understanding  the entropic interactions
stabilizing of micelles, membranes, soluble proteins, and hierarchical
biomolecular aggregation in aqueous solution has similarly advanced. It
is now recognized that scaled-particle theories can properly describe
primitive hydrophobic effects associated with the hydration of simple
mono-atomic species. Scaled-particle theories identify --- tentatively
at first, but firmly as further information accumulates --- a length
separating microscopic from macroscopic descriptions of hydration
structure. This establishes a radius at which microscopic and
macroscopic descriptions of hydration structure can be effectively
joined.   Together with primitive constitutive information specific to
liquid water, recognition of this joining radius provides an effective
description of hydrophobic effects for meso-scale aqueous structures.

This paper traces those advances, specifically by laying out the basic
view generalizing applications of scaled-particle approaches.  We
establish the micro-macroscopic joining length, discuss the length ---
analogous to the Tolman length --- associated with curvature corrections
of a surface area model of hydrophobic hydration free energies, and
finally examine the length-scales on which \emph{entropy convergence} of
hydration free energies are expected.

\subsection{Sketch of experimental characteristics of hydrophobic
hydration} Hydrophobic phenomena usually do not occur in isolation from
other interaction effects. The solutes that motivate study of
hydrophobic effects are typically molecularly complicated,
water-soluble, amphiphilic chain molecules.

Researchers studying these systems have been comfortable, however, with
a hydrophilic-hydrophobic dichotomy.  It is common to identify
contributions to the hydration free energy above and beyond obvious
hydrophilic interactions as hydrophobic effects \cite{Pratt:ECC}. This
is particularly true if the temperature dependences of the complementary
hydrophobic interactions are also distinctive
\cite{SpolarRS:HYDEIP,SpolarRS:COUOLF}. A helpful review of hydrophilic
electrostatic interactions involved in protein molecular structure with
an emphasis on the multiple length-scales involved appeared recently
\cite{SimonsonT:Eleadp}.   The present discussion emphasizes model
solutes, inert excluded volume models, that permit study of hydrophobic
effects exclusively.  Gases that are sparingly soluble in water, and
small hydrocarbon molecules, can be brought into correspondence with
hardcore molecular models. These models permit precision in isolating
the temperature signatures that are the target of studies of hydrophobic
effects.

High precision calorimetric studies show that unfavorable entropy
changes upon transferring nonpolar solutes into water dominate hydration
free energies at room temperature, and are only partly compensated by
favorable enthalpy changes. These measurements are incongruous with
\emph{regular solution} notions, which ascribe insolubility to
unfavorable cohesive interactions of the solute with water compared to
water with itself \cite{LazaridisT:Solsvc}. Of foremost importance, the
entropies and enthalpies of hydrophobic hydration are strongly
temperature dependent, following from the large positive heat capacity
changes observed. Thus, at elevated temperatures the roles of entropy
and enthalpy are reversed, with unfavorable enthalpies dominating
hydration free energies, partly compensated by favorable entropies. The
resulting solubilities of nonpolar gases are nonmonotonic, exhibiting a
solubility minimum between $T = $ 310~K and 350~K. Analogously, proteins
undergo hot \emph{and} cold denaturation
\cite{BRANDTSJF:THEOPD,Franks:91} as noted above, while ionic and
nonionic surfactants display a minimum in their critical micelle
concentrations with respect to temperature \cite{Chen:98,Chen:98b},
pointing to a common underlying mechanism with the solubility behavior
of nonpolar species. 

The experimental entropy changes for a range of hydrophobic solutes
intersect at values close to zero near
$T = $ 385~K
\cite{Privalov:79,Baldwin:86,PRIVALOVPL:STAOPA,Murphy:90,Makhatadze:95,
Lee:91,MULLERN:93}. The coincidence of this entropy convergence
temperature for hydrocarbons with comparable behavior in protein
unfolding has been used as empirical justification for the hydrophobic
core model for protein folding, and has influenced the interpretation of
biomolecular assembly. But in the complex context of soluble protein
molecules, a clear relevance of entropy convergence can be questioned
\cite{Robertson:97}.

It has been traditionally argued that these hydrophobic entropy effects
stem from orientational constraints on water molecules in the hydration
shell of nonpolar solutes, constraints that maintain the integrity of  a
hydrogen-bonding network forming cage-like structures or microscopic
\emph{icebergs} \cite{Frank:JCP:45}.  Theoretical studies of the impact
of local clathrate formation about krypton in water conclude that
literal clathrate structures are far too unfavorable to play a role in
hydration thermodynamics \cite{ashbaugh:03}. Experimental probes of the
local structure of water proximal to purely nonpolar solutes are scarce,
and hampered by the low solute-concentrations attainable. The structures
that have been studied by neutron and X-ray scattering techniques
suggest that while water surely adopts orientational preferences in the
hydration shell of nonpolar moieties, the solute-induced structure is
more disordered than  in ice or clathrate hydrates
\cite{BROADBENT:94,DeJong:97,Filipponi:97,Bowron:98,BowronDT:98B}.

Molecular-level investigations of hydrophobic hydration have been
largely theoretical and simulation efforts
\cite{Stillinger:73,Pierotti:76,HendersonJR:Solss,Pratt:02,HummerG:IT,
Hummer:98,Pohorille:90,Pratt:92}. Water structure in the vicinity of
hydrophobic species, including orientational preferences, has been
connected to the  entropies of hydrophobic hydration by a correlation
function expansion \cite{Lazaridis:92,Silverstein:2001,Ashbaugh:1996}.
Pratt and Pohorille, on the other hand, demonstrated that the low
solubility of atomic solutes in water arises from the narrow
distribution of cavity-opening fluctuations in water
\cite{Pratt:93,Pratt:92}.  With connections to the Pratt-Chandler theory
\cite{Pratt:77}  and its Gaussian field interpretation
\cite{Chandler:93,Lum:99}, an information theoretic (IT) model
\cite{Hummer:98,Hummer:00} then provided a quantitative link between the
microscopic density fluctuations determined from water oxygen pair
correlations and the hydration free energies of hard solutes. More
importantly, the IT model implicated the unusual equation-of-state
properties of water as a dominant factor in hydrophobic hydration,
differentiating water from other common solvents. 

In addition to capturing temperature and pressure effects associated
with hydrophobic hydration and interactions \cite{Hummer:98}, the IT
model provides a facile description of the enhanced solubility of
nonpolar species in D$_2$O compared to H$_2$O as a result of differences
in the isothermal compressibilities of these isotopic alternative forms
of water \cite{Hummer:00}. 

The latter point suggests again the length $\gamma \kappa_T$ of
Fig.~\ref{figEW}, because of the involvement of the measured isothermal
compressibility with its specific temperature dependence.  But the
product $\gamma \kappa_T$ involves the surface tension $\gamma$ also,
and the surface tension has been invoked in empirical correlations
connecting measurable properties of liquid water and hydrophobic
effects. Years ago, however, \cite{Tanford:79} pointed out
the large discrepancy between the measured water-hydrocarbon interfacial
tension and the effective microscopic surface tensions obtained from
hydrocarbon solubility data. A correspondence between macroscopic and
microscopic surface tensions has been contentious because of their
fundamentally different temperature dependence.  As noted above, it is
temperature dependences that have been the basic puzzles of hydrophobic
effects.

More recent examples of the distinction between molecular and
macroscopic hydrophobic interactions are found in measurements of the
long-range attractive force between macroscopic hydrophobic surfaces
\cite{Israelachvili:82,Pashley:85,CHRISTENSONHK:CAVATI}  which have not
been explained on the basis of molecular hydrophobic effects. Vibrational sum
frequency spectroscopy suggests that hydrogen-bonding of water molecules
is weaker at macroscopic water-carbon tetrachloride and water-hexane
interfaces than near individual hydrophobic species dissolved in water
\cite{Scatena:2001}.  The lack of a definitive interpretation of these
surface force measurements, and the changes in water energetics at
macroscopic interfaces underscores the need for a quantitative theory
beyond molecular hydrophobic effects. In general, the need for a
unified, quantitative description of both molecular and macroscopic
hydrophobic phenomena arises because hydrophobic driving forces play an
important role in self-assembly on multiple length-scales and the fact
that quantitative descriptions of these driving forces are derived from
molecular solubility data, macroscopic interfacial tension measurements,
or interpolations of these quantities
\cite{Ashbaugh:99,AshbaughHS:Effssa,Tanford:79,HERMANNRB:USEOSC,Sharp:91%
,GallicchioE:Entacd}.

\cite{Lum:99} suggested  bridging these disparate length-scales by
incorporating a Gaussian field theory for molecular level fluctuations
with mean-field theory for larger scale structures ultimately
responsible for macroscopic phase transitions. Their approach
successfully predicts many of the thermodynamic anomalies characteristic
of small molecule hydration, and goes further, predicting the onset of
long-range hydrophobic forces between surfaces as a result of an aqueous
liquid-vapor  phase transition in confined geometries. Indeed surface
force apparatus studies of the long-range hydrophobic interaction
\cite{CHRISTENSONHK:CAVATI} and simulations of water confined between
repulsive oblate ellipsoids observed cavitation between nonpolar
surfaces \cite{HuangX:Dewchp}, consistent with theoretical predictions.
Mean-field modeling and simulations of methane clusters, however,
suggest that when ubiquitous attractive interactions between water and
hydrophobic surfaces are taken into account, surface and confinement
induced local structural changes are suppressed
\cite{AshbaughHS:Effssa,Truskett:01,ChauPL:Comshh,%
DzubiellaJ:Redhab,HSA:2004}. Moreover, experiments on the effects of
electrolyte addition and degassing on the range of surface forces, and
the stability of surfactant free aqueous emulsions challenge the
theoretical predictions
\cite{ConsidineRF:Formbl,Kokkoli:89,PashleyRM:Effdfa,newPashley}.

A conceptual basis for unifying molecular and macroscopic hydrophobic
hydration can be found in  scaled-particle theory (SPT). Thirty years
ago \cite{Stillinger:73} presented an influential paper
on the application of the classic SPT of Reiss
\cite{Pierotti:76,Reiss:59,Reiss:65,Reiss:77} to the hydration
thermodynamics of purely excluded volume solutes. The purpose of that
paper was, in part, to illuminate the pitfalls and difficulties in
applying classic SPT, originally developed for  hard-sphere fluids, to
aqueous solvents \cite{BenNaim:67}. In doing so Stillinger opened up new
avenues of inquiry into hydrophobic hydration within the context of SPT.
Nevertheless, direct exploration of the validity and  consequences of
Stillinger's  revised theory have been rare \cite{Pratt:93,Pratt:92}. We
presently revisit SPT and critically discuss its implications in light
of our current understanding of hydrophobic hydration. For the first
time, we demonstrate that the revised SPT reproduces many of the
characteristic thermodynamic signatures of molecular hydrophobic effects
and can be used to extend the results of molecular simulations of small
hard hydrophobic  solutes in water to meso- and macroscopic surface
hydration. The present analysis provides insights into the differences
and similarities for hydrating molecular and macroscopic surfaces. In
addition, we examine the validity of surface area correlations commonly
used in biophysical models for hydration thermodynamics over a range of
length-scales, as well as the origins of entropy convergence behavior at
molecular lengths scales and how solute size moderates the convergence
temperature.

\vfil

\section{A primer on scaled-particle theory}

\subsection{Classic scaled-particle theory}\label{sec:classic}
The chemical potential of a hydrated mono-atomic solute, `A', can be expressed
formally as
\begin{eqnarray} \mu_\mathrm{A}\left(\mathrm{aq}\right) = kT
\ln\left\lbrack\rho_\mathrm{A}\left(\mathrm{aq}\right)\Lambda_\mathrm{A}{}^3\right\rbrack +
\mu_\mathrm{A}^\mathrm{ex} \left(\mathrm{aq}\right)
\end{eqnarray} 
where we have adopted Ben-Naim's standard state in the definition of the
chemical potential \cite{BENNAIMA:SOLTON}.  Here $k$ is Boltzmann's
constant, $T$  is the temperature,
$\rho_\mathrm{A}\left(\mathrm{aq}\right)$ is the solute number density
in the solution, $\Lambda_\mathrm{A}$  is the thermal de Broglie
wavelength of the solute, and $\mu_\mathrm{A}^\mathrm{ex}$ is the excess
chemical potential, \emph{i.e.}, the coupling work of turning on
interactions between the solute and water, which disappears if those
interactions vanish. At thermal equilibrium the Ostwald partition
coefficient determining the distribution of the solute between an
aqueous and ideal gas [$\mu_\mathrm{A}^\mathrm{ex}
\left(\mathrm{ideal}\right)$ = 0] phase at infinite dilution is
\cite{Pollack:91}
\begin{eqnarray}
K_\mathrm{eq} =  \frac{\rho_\mathrm{A}\left(\mathrm{aq}\right)}{\rho_\mathrm{A}\left(\mathrm{ideal}\right)} = \exp\left\lbrack -
\mu_\mathrm{A}^\mathrm{ex} \left(\mathrm{aq}\right) /kT\right\rbrack.
\label{Ostwald}
\end{eqnarray}
Thus, the excess chemical potential is central to resolving the aqueous
solubility of the solute. Confining our discussion to impenetrable hard
sphere (HS) solutes, the solute excess chemical potential in water is
\begin{eqnarray}
\mu_\mathrm{A}^\mathrm{ex} \left(\mathrm{aq}\right) = - kT \ln p_0(R)
\end{eqnarray}
where the  insertion probability, $p_0(R)$, is the probability that a
solute-sized stencil randomly placed in water is devoid of water oxygen
centers;  this  follows directly from Widom's potential distribution
theorem for species interacting with a hard potential
\cite{Widom:JPC:82,Pratt:99}. The solvent accessible radius, $R$, is the
radius of closest approach between the solute center and a water oxygen.
The solvent accessible radius is the sum of the van der Waals
radius of a water molecule and the radius of the hard-sphere solute,
\emph{i.e.,} $R$  = $\left(\sigma_\mathrm{WW}
+\sigma_\mathrm{AA}\right)/2$. The insertion probability is the
fractional free volume offered by the solution --- or the
\emph{available} volume in a more specialized language attributed to
Boltzmann \cite{Stell:MM:85}, \emph{i.e.}, $p_0(R) $ =
$V_\mathrm{free}/V_\mathrm{total}$.

\begin{figure}[h]
\includegraphics[scale=0.35]{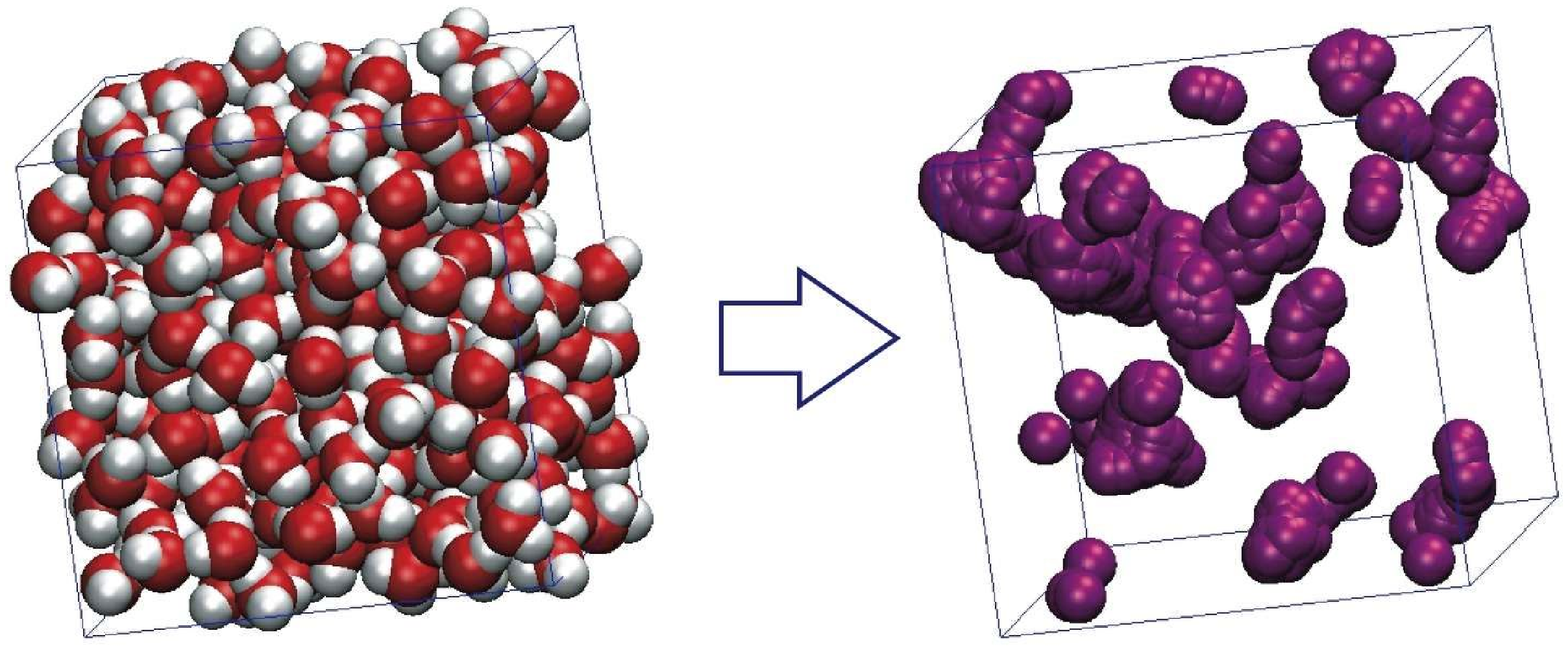}
\caption{The box on the left shows a configuration of water molecules
taken from a simulation of liquid water  at the density of the  liquid
in  coexistence with vapor at 300 K.  The box on the right shows hard
spheres of diameter 2.8~\AA\ that can be successfully placed into the
configuration on the right without overlap of the van der Waals volume
of the water molecules.  The insertion probability $p_0$ is determined
as the volume accessible to centers of the purple spheres divided by the
geometric volume of the box.}\label{fig13}
\end{figure}

Figure~\ref{fig13} gives a physical picture of this available volume.
Given a molecular snapshot of liquid water --- the left  box in
Fig.~\ref{fig13}  --- $V_\mathrm{free}$ is determined by the points at
which a hard sphere could be successfully implanted.  Graphic display of
those successful placements thus yields a negative image --- the right box
in Fig.~\ref{fig13} --- of the molecular configuration from which the
analysis started. $V_\mathrm{free}$ decreases with increasing cavity
radius,  as successful insertions become increasingly rare. $p_0(R)$ is
subsequently determined as an ensemble average over a large number of
molecular configurations.

An alternative relationship for the chemical potential that draws a
connection to the structure of water in the vicinity of the solute is
\begin{eqnarray}
\mu_\mathrm{A}^\mathrm{ex} \left(\mathrm{aq}\right) 
= kT \int_0^R \rho_\mathrm{W} G(r)4 \pi r^2\dif r
\label{eq:scaling_integral}
\end{eqnarray}
where $\rho_\mathrm{W} G(r) $ is the density of water in contact with a
hard solute of contact radius $r$ \cite{Stillinger:73}. The product
$kT  \rho_\mathrm{W} G(r) $ has units of force/area, and is the
pressure due to solvent collisions with the hard  surface of the solute. The
chemical potential then is the reversible pressure-volume work required
to expand the solute into the solution. The etymology of scaled-particle
theory derives from this expression \cite{Reiss:65,Reiss:77} since
the solute is introduced by scaling-up from a particle with a solvent
accessible radius of zero to a final size of $R$. The relationship
between the contact function and the  insertion probability is
\begin{eqnarray}
G(R) =- \frac{1}{4\pi R^2\rho_\mathrm{W}}\frac{\partial\ln p_0(R)}{\partial R}~,
\label{G-def}
\end{eqnarray}
as determined by differentiation of Eq.~\eqref{eq:scaling_integral}. For
sufficiently small solute cavities  that only one solvent molecule
might fit within the cavity boundary, the  insertion probability
is 
\begin{eqnarray}
p_0(R\rightarrow 0) = 1-\frac{4\pi}{3} \rho_\mathrm{W} R^3~.
\label{small-p}
\end{eqnarray} 
This is merely $V_\mathrm{free}/V_\mathrm{total}$ in the case that the
solute is so small that it never interacts with more than one solvent
molecule exclusion sphere at once.  Then the excluded volume is $
\left(4\pi R^3/3\right)N_\mathrm{W}$; the more general case comes-up
below. The expression corresponding to Eq.~\ref{small-p} for the contact
function is
\begin{eqnarray} 
G(R\rightarrow 0) = \frac{1}{1-\frac{4\pi}{3} \rho_\mathrm{W} R^3}~.
\end{eqnarray}
These expressions are accurate in solvents with realistic interactions
for cavities up to a radius of $R\approx \sigma_\mathrm{WW}/2$, and  are
exact in hard sphere fluids then because the solute corresponds to a
point particle with a van der Waals radius of zero.  Larger hard-sphere
solutes can contact more than  one solvent molecule at once, and
addressing larger solutes  requires information on dauntingly
complicated multi-body correlations.

In the limit of a macroscopically large cavity, the contact correlation
function can be represented as an asymptotic expansion in $1/R$
\begin{eqnarray}
G(R) \sim\sum_{j\ge 0} \frac{G_j}{R^j}~.
\label{asymp}
\end{eqnarray}
Retaining contributions up to $j$ = 2 yields an expression equivalent in
form to that required by classical thermodynamics for the force acting
on the cavity surface
\cite{Stillinger:73,Pierotti:76,Reiss:65,Reiss:77,HendersonJR:Solss}
\begin{eqnarray}
kT\rho_\mathrm{W} G(R) \sim p + \frac{2\gamma_\infty}{R}  - \frac{4\gamma_\infty\delta}{R^2}
\label{eq:tolman}
\end{eqnarray}
\begin{minipage}{3.4in}
\bigskip
\noindent where $p$ is here to equal the liquid saturation pressure
$p_\mathrm{sat}$, $\gamma_\infty$ is henceforth the surface tension of a
flat interface that was denoted by $\gamma$ above, and $\delta$
\cite{TOLMANRC:THEEOD}  describes the initial curvature correction to
the surface tension.\footnote{The notation of Eq.~\eqref{eq:tolman}
follows a  typographic confusion wide-spread across the present problem.
$\delta$ here is one-half the conventional Tolman length.  Expressed
more basically the final term on Eq.~\eqref{eq:tolman} is
$(2\gamma_\infty/R)$ {\em times} the Tolman length. See
\cite{MoodyMP:Curdst,HendersonJR:Solss}.  We thank J. R. Henderson for a
discussion of this point.} The coefficient $G_3$ is zero so that the
chemical potential is free of logarithmic contributions
\cite{STILLING*FH:FREEIP} though analyses of the possibility of
logarithmic corrections are still of interest
\cite{EvansR:WetcsN,EvansR:Nonccs}. Higher order terms in the asymptotic
expansion are not generally available. Such considerations motivated
\cite{Reiss:59} to truncate Eq.~\eqref{asymp} after the initial
curvature correction in order to develop a tractable, physically
reasonable model for the contact function.
\end{minipage}

Evaluating the $j$ = 0  term in Eq.~\eqref{asymp} with the measured
equation of state, and the $j$ = 1, 2 terms in the expansion by
requiring the microscopic and macroscopic limits smoothly meet at
$\sigma_\mathrm{WW}/2$, yields
\begin{widetext}
\begin{eqnarray}
G(R) = \left\{ 
\begin{array}{c@{\quad}l}
\frac{1}{1-\frac{4\pi}{3}\rho_\mathrm{W} R^3}~, &  R\le \sigma_\mathrm{WW}/2 \\
\frac{p_\mathrm{sat}}{kT\rho_\mathrm{W}}~  
+ \left\lbrack \frac{2+\eta}{\left(1-\eta\right)^2} -
\frac{2p_\mathrm{sat}}{kT\rho_\mathrm{W}}\right\rbrack\left(\frac{\sigma_\mathrm{WW}}{2R}\right) \\
+ \left\lbrack -
\frac{\left(1+2\eta\right)}{\left(1-\eta\right)^2}+\frac{p_\mathrm{sat}}{kT\rho_\mathrm{W}}\right\rbrack\left(\frac{\sigma_\mathrm{WW}}{2R}\right)^2~, & 
R>\sigma_\mathrm{WW}/2
\end{array}
\right.
\label{spt-G}
\end{eqnarray}
where  $\eta$ = $\frac{\pi}{6}\rho_\mathrm{W}
\sigma_\mathrm{WW}{}^3$ is the solvent packing fraction. Integration of
the contact function yields the  excess chemical potential
\begin{eqnarray}
\frac{\mu_\mathrm{A}^\mathrm{ex}}{kT} = \left\{ 
\begin{array}{c@{\quad}r}
  - \ln\left(1-\frac{4\pi}{3}\rho_\mathrm{W} R^3\right)~, &  R\le \sigma_\mathrm{WW}/2 \\
\left\lbrack -\ln\left(1-\eta\right) + \frac{9\eta^2}{2\left(1-\eta\right)^2}
- \frac{\eta p_\mathrm{sat}}{kT\rho_\mathrm{W}}\right\rbrack 
 + \left\lbrack -
 \frac{3\eta\left(1+2\eta\right)}{\left(1-\eta\right)^2} +
 \frac{3\eta p_\mathrm{sat}}{kT\rho_\mathrm{W}}\right\rbrack
 \left(\frac{2R}{\sigma_\mathrm{WW}}\right) &  \\
 + \left\lbrack \frac{3\eta\left(2 +
 \eta\right)}{2\left(1-\eta\right)^2}-\frac{3\eta p_\mathrm{sat}}{kT\rho_\mathrm{W}}\right\rbrack
 \left(\frac{2R}{\sigma_\mathrm{WW}}\right)^2 
 + \frac{\eta p_\mathrm{sat}}{kT\rho_\mathrm{W}}\left(\frac{2R}{\sigma_\mathrm{WW}}\right)^3~, & R>\sigma_\mathrm{WW}/2~.
\end{array}
\right.
\label{spt-mu}
\end{eqnarray}
\end{widetext}
Eqs.~\eqref{spt-G} and \eqref{spt-mu} constitute the classic SPT
originally developed for  the hard sphere fluid
\cite{Reiss:65,Reiss:77}, but which was subsequently applied to water by
\cite{Pierotti:76} and \cite{Lee:85}. Fundamental difficulties arise in
the application of classic SPT to water, however, including the
erroneous prediction that the surface tension of water increases with
temperature and passes through a maximum near $T = $ 425~K
\cite{Stillinger:73}.

\subsection{Revised scaled-particle theory} 
The scaled-particle model described above incorporates little molecular
detail beyond the assigned van der Waals diameter, $\sigma_\mathrm{WW}$,
that might differentiate water from other solvents, and thereby limits
the interpretation of complex hydration phenomena. To consider this
problem more generally, we note that the insertion probability is
formally expressed on the basis of solvent structure by an
inclusion-exclusion development \cite{Reiss:65,Reiss:77,Stillinger:73}
\begin{widetext}
\begin{eqnarray}
p_0\left( R\right) = 1 + 
\sum_{n\ge
1}\frac{\left(-\rho_\mathrm{W}\right)^n}{n!}\int_{V(R)}\ldots\int_{V(R)}
g^{(n)}\left(\vec{r}_1 \ldots\vec{r}_n\right)\dif^3 r_1\ldots \dif^3 r_n
\label{in-ex}
\end{eqnarray}
where $V(R)$ = $4\pi R^3/3$ is the observation volume, and
$g^{(n)}\left(\vec{r}_1 \ldots\vec{r}_n\right)$ are the $n$-body solvent
oxygen distribution functions. The terms in this series vanish for $n$
exceeding the maximum number of solvent molecular centers that can be
packed into a sphere of volume $4 \pi R^3 /3$. It is on this basis that
the limiting result, Eq.~\eqref{small-p}, is established.  As noted
above, these distribution functions are complicated, not routinely
available beyond the pair-distribution function ($n$ = 2), and there
has been only one detailed investigation of terms beyond 
2nd order  \cite{GomezMA:Molrdm,Pratt:99}. Considering the small cavity
pair-correlation ($n$ = 2) contribution and the asymptotic macroscopic
thermodynamic limits, Stillinger proposed a revised expression for the
cavity contact function \cite{Stillinger:73}
\begin{eqnarray}
G(R) = \left\{ 
\begin{array}{c@{\quad}l}
\frac{1+\frac{\pi\rho_\mathrm{W}}{R}\int_{0}^{2R}
g^{(2)}(r)r^2\left(r-2R\right)\dif r}{1-\frac{4\pi}{3} \rho_\mathrm{W} R^3 +
\left(\frac{\pi\rho_\mathrm{W}}{R}\right)^2\int_{0}^{2R}
g^{(2)}(r)\left(r^3/6-2R^2r + 8R^3/3)\right)\dif r}~, &  R\le R^\ast \\
\frac{p_\mathrm{sat}}{kT\rho_\mathrm{W}} + \frac{2\gamma_\infty}{kT\rho_\mathrm{W}R}  -
\frac{4\gamma_\infty\delta}{kT\rho_\mathrm{W}R^2}+\frac{\lambda}{R^4}~, &
R>R^\ast
\end{array}
\right.
\label{rspt-G}
\end{eqnarray}
\end{widetext}
where $R^\ast$ is the radius at which $n$ = 3 correlations begin to
contribute to the cavity insertion probability of Eq.~\ref{in-ex}. While
the experimental pressure, surface tension, density, and solvent radial
distribution function are employed, $\delta$ and $\lambda$ are treated
as adjustable parameters chosen so that the small cavity and macroscopic
limits of the contact function join smoothly at $R^\ast$. This
expression incorporates molecular information on the pair structure of
water as well as the known macroscopic properties of bulk water and its
interfacial behavior, and thereby is expected to discriminate more
sensitively between water \cite{AshbaughHS:Effssa} and other solvents
\cite{Huang:00}. Indeed, it has been demonstrated by extensive molecular
simulations that Eq.~\eqref{rspt-G} provides a  description superior to
the classic SPT expression Eq.~\eqref{spt-G} of the solvent contact
density for solutes several times larger than the solvent.

Stillinger's revised SPT prediction for $G(R)$ relies on the assumption
that multi-body water correlations at intermediate, but molecule-sized,
solute radii are adequately represented by the parameters $\delta$ and
$\lambda$ fitted at a  radius $R^\ast$. Numerical experimentation
\cite{Pratt:92} with this parameterization shows that the  revised SPT
is most sensitive to the parameter $R^\ast$. This parameterization might
be improved by involving results over a range of radii, including solute
sizes for which multi-body correlations are significant. While this
information is not readily available experimentally, multi-body
correlation contributions to the hard-sphere solute chemical potential can be
interrogated by direct evaluation of the insertion probabilities from
molecular simulations of water. In the spirit of Stillinger's revised
SPT, we interpolate between the chemical potential evaluated for molecular length-scales
from simulation, and the macroscopic thermodynamic limit
\begin{multline}
\mu_\mathrm{A}^\mathrm{ex}(R) = - kT\ln p_0(R)\vert_\mathrm{sim}f(R)  \\
 + \mu_\mathrm{A}^\mathrm{ex}(R)\vert_\mathrm{macro}\left[1-f(R)\right]
 \label{rrspt-mu.0}
\end{multline}
where $f(R)$ is a switching function equal to one below $R_\mathrm{sim}$
and zero above $R_\mathrm{macro}$, smoothly interpolating between these
between these two limits. Presently we use a cubic spline interpolating
function, though other reasonable functions yield essentially
indistinguishable predictions. The macroscopic  chemical potential,
determined by integration of the macroscopic cavity expression
Eq.~\eqref{rspt-G} in
\begin{multline}
\mu_\mathrm{A}^\mathrm{ex}(R)\vert_\mathrm{macro} =  - \frac{4\pi kT\rho_\mathrm{W} \lambda}{R} + \epsilon 
- 16\pi R\gamma_\infty \delta \\
 + 4 \pi R^2\gamma_\infty + \frac{4\pi}{3}R^3 p_\mathrm{sat}
 \label{rrspt-mu}
\end{multline}
Rather than fitting the microscopic and macroscopic limits at a single
point as in Eq.~\eqref{rspt-G}, the parameters $\delta$, $\lambda$, and
the integration constant $\epsilon$  are fitted to the simulation results
between $R_\mathrm{sim}$ and $R_\mathrm{macro}$. The  contact correlation
function is then determined by differentiation of the chemical potential
[Eq.~\eqref{G-def}]. An additional benefit of fitting Eq.~\eqref{rrspt-mu} 
to the simulation insertion probabilities is that we do not have to
evaluate numerically first and second derivatives of the simulated
insertion probabilities, which become more statistically uncertain with
increasing cavity size.

\subsection{Computational implementation}  The computational
implementation follows standard procedures for sampling molecular
configurations of liquid water, and builds from the original work of
\cite{Pohorille:90,Pratt:92,Pratt:93} in evaluating cavity statistics
therefrom. Water configurations were generated using Monte Carlo
simulations in the canonical ensemble \cite{Frenkel-Smit}. Bulk water
was modeled using 268 SPC/E water molecules with periodic boundary
conditions \cite{BERENDSENHJC:THEMTI}. SPC/E was chosen because it
provides an accurate representations  of the structure,
equation-of-state, and interfacial tension of liquid water over a broad
range of temperatures \cite{HuraG:03,Alejandre:95}. Lennard-Jones
potential interactions were evaluated smoothly truncating the potential
based on the separation of water oxygen atoms between 9.5~\AA\ and
10~\AA, while longer ranged electrostatic interactions were calculated
using Ewald summation with conducting boundary conditions
\cite{Frenkel-Smit}. Simulations were carried out from $T = $ 260~K to
470~K in 10~K increments at the experimental liquid density along the
saturation curve and into the supercooled regime \cite{HAREDE:DENOSH}.
After an equilibration phase of at least $10^5$ MC passes (where one
pass corresponds to one attempted move per water molecule with 30\% move
acceptance), $5\times 10^6$ MC production passes were carried out for
analysis of thermodynamic averages. After each 50 MC passes, $10^5$
particle insertions were attempted to estimate $p_0(R)$, so a total of
$10^{10}$ insertions were attempted at each temperature. Statistical
uncertainties were determined by grouping results into block averages
over $10^6$ MC passes each.

\begin{figure}
\includegraphics[scale=0.55]{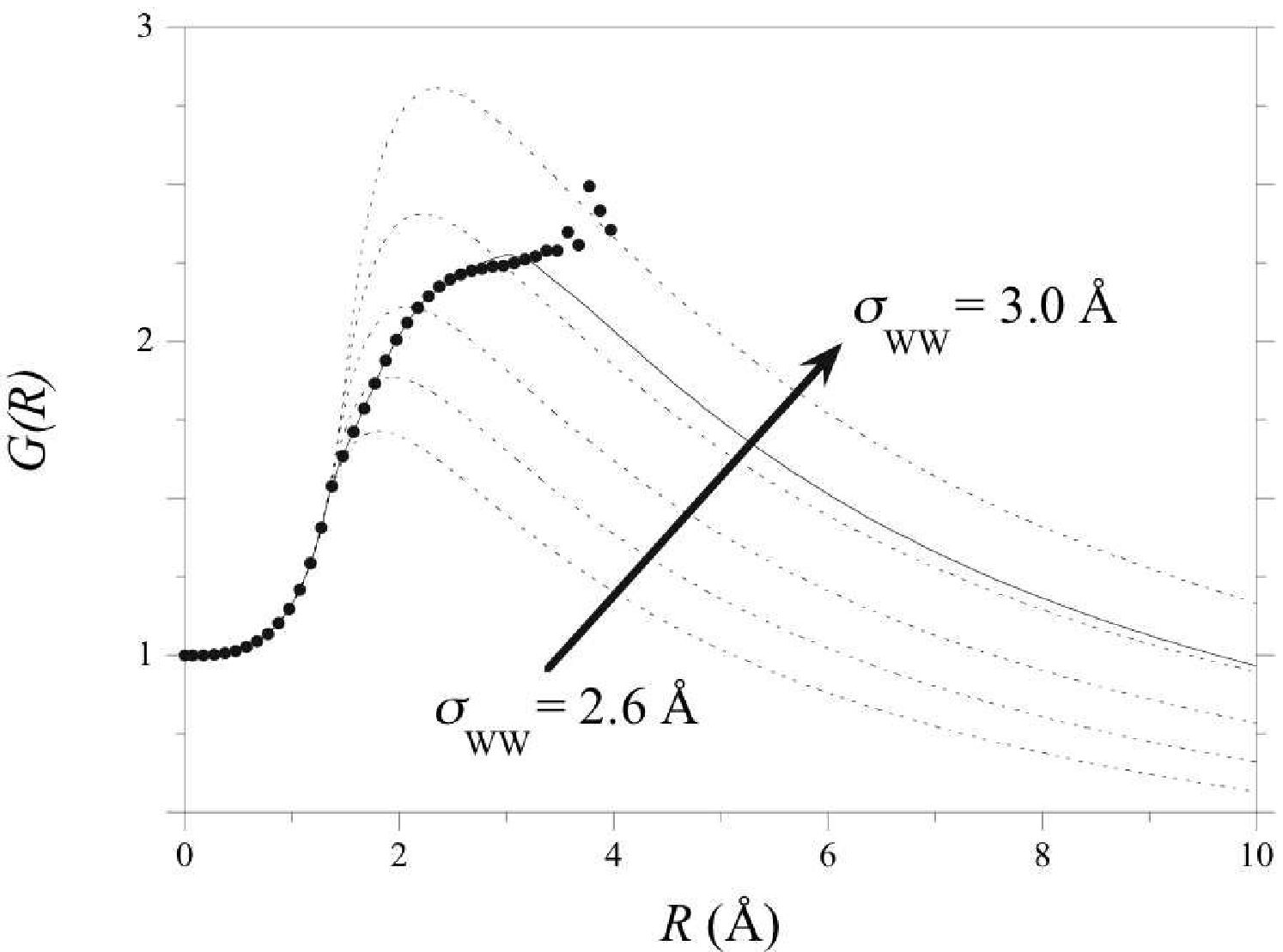}
\caption{Cavity contact function for water at $ T = $ 300~K
at the liquid saturation conditions. The points are obtained
by differentiation of the simulation cavity insertion
probabilities.  The dashed lines are obtained from Reiss's
original SPT theory predictions for a hard sphere solvent,
Eq.~\eqref{spt-G}, using effective water hard sphere
diameters between $\sigma_\mathrm{WW}$ = 2.6~\AA\ to
3.0~\AA\ in 0.1~\AA\ increments. The solid line is obtained
by differentiation of the revised SPT, Eq.~\eqref{rrspt-mu},
fitted to the simulation results between $R_\mathrm{sim}$ =
2.5~\AA\ and $R_\mathrm{macro}$ = 3.5~\AA.}\label{fig1}
\end{figure} 

\section{Application to hydrophobic hydration} 
\subsection{Cavity contact functions: the micro-macro joining boundary}
\begin{figure} 
\includegraphics[scale=0.5]{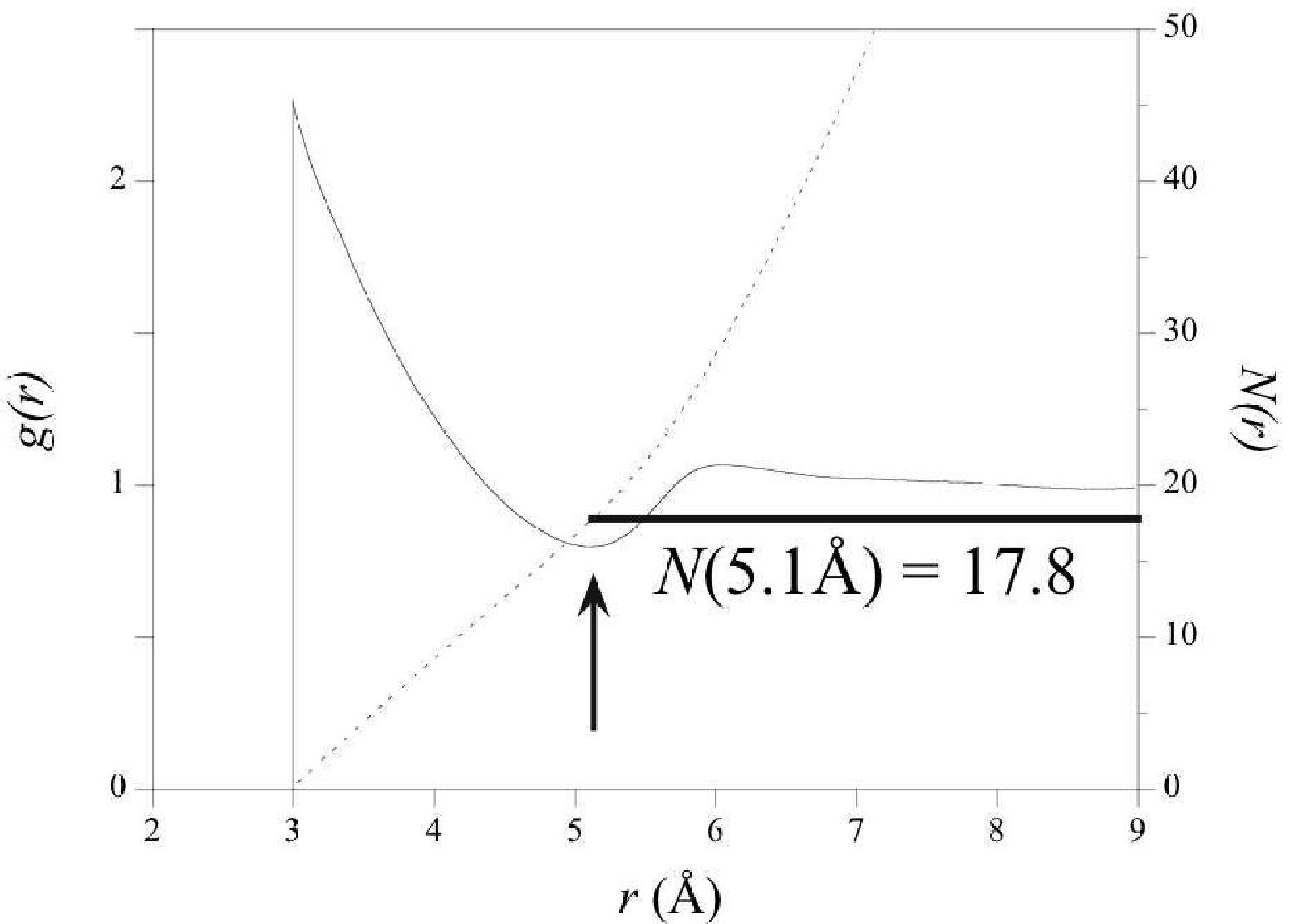}
\caption{Cavity-water oxygen radial distribution function for a 
3~\AA\ cavity at $T = $ 300~K. The thin solid line indicates the radial
distribution function, while the dashed line indicates the radial
integral $N(r) = \int_0^r\rho_\mathrm{W}g(\lambda)4\pi\lambda^2d\lambda$. The
occupation of the first hydration shell, corresponding to the first
minimum in $g(r)$ at 5.1~\AA, is 17.8 water molecules as indicated by
the thick horizontal line.  Note that the first minimum, which
physically discriminates between first and succeeding hydration shells,
is mild, and structuring of outer hydration shells is weak
\cite{Pratt:93}.  These features are in good qualitative agreement with
the predictions of the Pratt-Chandler theory \cite{Pratt:77}, though
that specific theory has been substantially amended
\cite{Pratt:02}.\label{fig11}\vfill}
\end{figure}

The cavity contact function at $T = $ 300~K is shown in Fig.~\ref{fig1}.
Beginning at a value of one at zero radius, the cavity contact density
increases with increasing $R$. Simulation values of $G(R)$ appear to
plateau at a maximum near 3~\AA.   Just beyond this radius, the
simulation results for $G(R)$ become progressively noisy as a result of
poor sampling of infrequent large cavity fluctuations. Detailed
calculations for specific values of $R$ greater than 3~\AA\
\cite{AshbaughHS:Effssa} have established that this is indeed the region
of a maximum in $G(R)$, and that $G(R)$ is qualitatively described by
Stillinger's revised scale particle model.  A dominating observation is
that this curve imposes a non-arbitrary definition of a length-scale for
the present problem: the radius $R_\mathrm{max}$ at which $G(R)$ is a
maximum. Solutes with smaller radii are identified as intrinsically
microscopic in scale. The description of larger solutes can be built
from a macroscopic perspective. An interpolative strategy extending to
large solutes, such as that adopted here, is likely to be  effective if
the region at which the molecular and macroscopic expressions are joined
encompasses $R_\mathrm{max}$. The revised SPT fit, determined by
differentiation of Eq.~\eqref{rrspt-mu.0} fitted to the simulation
insertion results between $R_\mathrm{sim}$ = 2.5~\AA\ and
$R_\mathrm{macro}$ = 3.5~\AA, extends $G(R)$ to $R$ larger than observed
directly. The revised SPT result places the maximum contact density at
$R$ = 3.0~\AA\ where $G(R)$ $\approx$ 2.3. Solutes of this size are
candidates for \emph{most hydrophobic} because the compressive pressure
exerted by the solvent is largest in this case; see Fig.~\ref{fig11}.

Beyond this maximum, water pulls away from the cavity surface with
increasing size.  At the size $R \approx$ 10~\AA\ the contact density
equals the bulk density of water, decreasing  further for larger
cavities. In the limit $R\rightarrow\infty$, the contact correlation
approaches $\frac{p_\mathrm{sat}}{\rho_\mathrm{W}kT}\approx 2
\times10^{-5}$ for water at $T = $ 300~K, and the pressures here are
sufficiently low that they do not influence the contact  function for
molecular and mesoscopic cavities at any of the temperatures considered.

This apparently anomalous drying behavior was anticipated by
\cite{Stillinger:73},  and has only recently been confirmed by molecular
simulations in Lennard-Jones and aqueous solvents
\cite{AshbaughHS:Effssa,Huang:00}.  The surface drying has been
previously interpreted in terms of an effective expulsion potential
between water and the solute cavity \cite{Hummer:98b,Weeks:98}.  In bulk
water, the individual water molecules feel attractive interactions with
the other water molecules, and  the average force on a water molecule in
the bulk is zero. To approach a large solute, however, a water molecule
must shed hydration partners and/or limit their possible configurations.
This unbalances the interactions with the aqueous medium and gives rise
to an additional repulsive force between a water molecule and the
surface. If the solute is unable to compensate for these lost
interactions to counter the cavity expulsion potential, water is
repelled by the surface.

The classic SPT prediction for the contact function [Eq.~\eqref{spt-G}]
is in qualitative agreement with the simulation and revised SPT results
[Fig.~\ref{fig1}]. Classic SPT predicts a maximum in the contact
density, followed by a decrease to values below the bulk density of
water with increasing cavity radius. The quantitative agreement is poor,
however, even if the effective diameter, $\sigma_\mathrm{WW}$, of water
is treated as an adjustable parameter. Notably, classic SPT predicts
that the maximum in $G(R)$ is shifted to smaller radii of $R \approx$
2~\AA. Water has an open structure favoring larger cavities at these
packing fractions \cite{Pohorille:90,Pratt:92}. The resulting maximum in
the pressure acting on the solute surface, $kT \rho_\mathrm{W} G(R)$,
then is shifted out to larger cavity radii for water.

If the objective of classic SPT is to reproduce the chemical potentials
of solutes using Eq.~\eqref{spt-mu} up to radii of $R \approx$ 3.3~\AA,
encompassing the sizes of a number of nonpolar gases, a typical
diameter, $\sigma_\mathrm{WW}$, of water fitted to experimental data is
2.7~\AA\ \cite{Lee:85}, though a more appropriate value based on the
simulation results reported in Fig.~\ref{fig1} is 2.8~\AA. In this case,
the fitted radius of water splits the difference between the
over-prediction of $G(R)$ at intermediate radii by the classic SPT, and
the under-prediction of $G(R)$ at radii close to the maximum solute size
to balance out these inaccuracies in the calculation of the chemical
potential. This $\sigma_\mathrm{WW}$ assigned to water molecules then is
a consequence of the fitting, and does not contribute to the
interpretation of the molecular signatures of hydrophobicity. Indeed, if
we extend the predictions of classic SPT outside the range fitted for
small solutes, we find the theory under-predicts the hydration free
energies of mesoscopic cavities, and that a larger water diameter ---
$\sigma_\mathrm{WW}\approx $ 2.9~\AA! --- is required to match the drying
observed in $G(R)$ [Fig.~\ref{fig1}]. 


While the variation of this size parameter may seem small, the chemical
potential depends on the integral of Eq.~\ref{eq:scaling_integral}, and
small differences in $\sigma_\mathrm{WW}$ significantly alter the
predictions. Broadly viewed, this is the natural observation that slight
adjustment of boundary information in boundary value  problems can make
large changes away from the boundary.

\begin{figure} 
\includegraphics[scale=0.5]{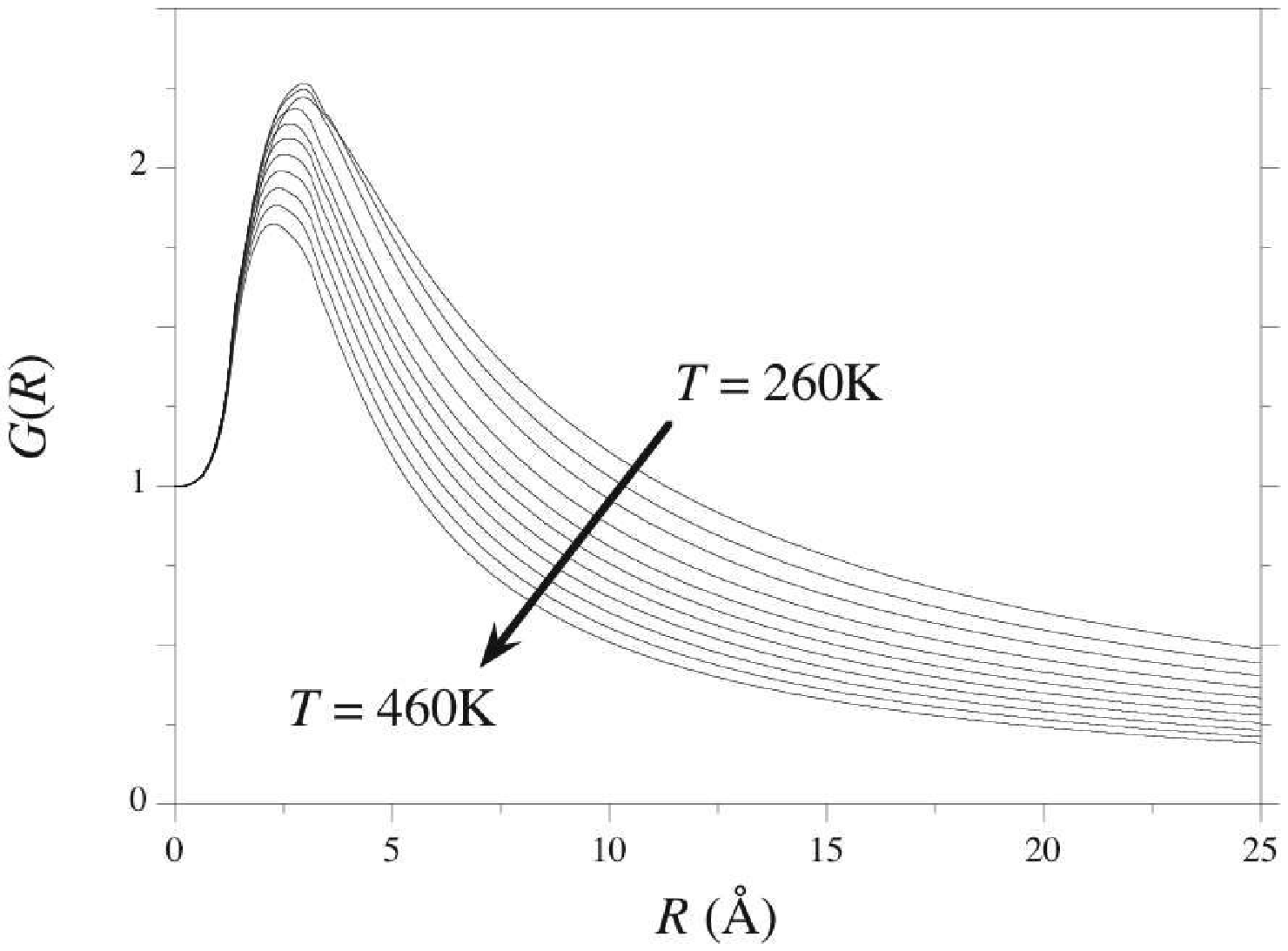}
\caption{Cavity contact  function for water as a function of temperature
along the liquid saturation curve determined by the revised SPT,
Eq.~\eqref{rrspt-mu.0}, with $R_\mathrm{sim}$ = 2.5~\AA\ and
$R_\mathrm{macro}$ = 3.5~\AA.  Results are shown between $T = $ 260~K to
460~K in 20~K increments. Notice that the length defined by the maximum
of this curve \emph{decreases} with decreasing density following
increasing temperature along the saturation curve.  This is consistent
with the suggestion of Fig.~\ref{figEW}.
\label{fig2}}
\end{figure}

Revised SPT predictions for $G(R)$ as a function of temperature are
shown in Fig.~\ref{fig2}. While all the curves are qualitatively
similar, the magnitude of the contact function decreases with increasing
temperature. Classic SPT fails to describe this temperature dependence
of $G(R)$.  Fig.~\ref{fig2} also indicates that the length defined by
the maximum of this curve \emph{decreases} with increasing temperature,
consistent with the suggestion of Fig.~\ref{figEW}. In following
sections we use these revised SPT results to draw conclusions regarding
the size and temperature dependence of the hydration of hydrophobic hard
spheres.

The discussion in Sec.~\ref{EWsubsection} of the length-scale
$\gamma_\infty \kappa_T$ sheds some light on  the results of
Fig.~\ref{fig2}. Specifically, we can put together the crudest of 
models of the hydration free energies for the small-scale and
large-scale problems, and in that way get a crude characterization of
the length at which these different descriptions for
$G\left(\lambda\right)$ match.  For the small-scale --- but not the
smallest-scale as in  Eq.~\ref{small-p}, the information theory approach
suggested \cite{Garde:96,Pratt:02}
$\beta\mu_\mathrm{A}^\mathrm{ex}\approx \left\langle
n\right\rangle_0{}^2/2 \left\langle \delta n^2\right\rangle_0$. If we
evaluate everything on a macroscopic basis, then $G\left(R\right) 
\approx \frac{\beta}{2\rho_\mathrm{W} \kappa_T}$ from Eq.~\ref{G-def} .
This is an extremely crude estimate, but we rely only the trends with
changes in temperature in the  present argument. For the large-scale
problem, we use $\beta\mu_\mathrm{A}^\mathrm{ex}\approx 4\pi
R^2\beta\gamma_\infty$, expecting that the  pressure will be negligibly
low for this consideration.  Then $G\left(R\right)  \approx
\frac{\beta\gamma_\infty}{R\rho_\mathrm{W}}$.  These two estimates match
when $R_\mathrm{max} \approx 2\gamma_\infty\kappa_T$.  The first of
these estimates, that from the information theory contribution evaluated
on macroscopic information, is too large by about a factor of 4. 
Exploiting an empirical factor of 4 would bring the estimate
$R_\mathrm{max} \approx 4\times 2\gamma_\infty\kappa_T$ into a realistic
range of molecular sizes. But the present argument is crude enough that
only the dimensionally natural linear correlation with  the length
$\gamma_\infty \kappa_T$ should be taken seriously.

\subsection{Physical relevance of hard-core model solutes to structural
theories of hydrophobic effects} Hard core models of solute-water
interactions studied in this context, which are the basis of
scaled-particle theories, are motivated \cite{Pratt:02} by the broad
success of van~der~Waals theories of liquids
\cite{BARKERJA:WHAIL-,WCA,Widom:67,LebowitzJL:STASFB}. Physically
expressed, van~der~Waals approaches are appropriate for the description
of dense liquids because the disorder and high density can limit
structural fluctuations to length-scales associated with the variation
of intermolecular repulsive interactions, small compared to the range of
intermolecular attractive interactions.  When attractive intermolecular
interactions are weak on a thermal scale, that helps too since a
van~der~Waal approach treats those interactions perturbatively. Then an
approach which separates the effects of intermolecular repulsive and
attractive interactions can be successful.  In that case, repulsive
interactions present the first challenge to theories, and hard-core
interactions are natural simple models for those excluded volume
interactions.  This is the argument for the  physical relevance of
hard-core models solutes to theories of hydrophobic effects.

With this background, the most  important {\em  physical} observation on
the large $R$ behavior of the $G(R)$ results of Fig.~\ref{fig2} is that
those results would be expected to be sensitive to attractive
solute-water interactions, if they were to be included.  When $R$ is
large, the local density in the vicinity of the hard sphere solute can
be low, and the argument above that fluctuations do not access the
length-scales comparable to the range of natural attractive interactions
does not apply.  Simulation evidence does support the view that
results can be sensitive to inclusion of natural attractive interactions
when the solution structures have length scales substantially greater
than  $R_\mathrm{max}$
\cite{HummerG:Watcth,AshbaughHS:Effssa,Truskett:01,ChauPL:Comshh,%
DzubiellaJ:Redhab,vanSwolF:WETAAF,HENDERSONJR:GRAPDO,WallqvistA:Amsd,%
ZhouRH:Hydcmp}.

Nevertheless, hard-core models of solute-water interactions serve as an
valuable reference point for at least two reasons.  A first reason is
conceptual and reductionist.  This simplified case has historically been
considered as expressing the basic puzzle of hydrophobic effects.  (The
extent to which that is true is one of the issues addressed here!) 
Solving this basic puzzle enables specific cases to be described by
combination of what is understood for the simpler cases. A second reason
that hard-core models of solute-water interactions are valuable is that
for $R$ not too large the results should be less sensitive specifically
to  the case of physical interest. (Support for this view is noted at
the appropriate places in the succeeding discussions.) From this point
of view, then, the careful study  of the large $R$ behavior of hard
sphere $G(R)$ assists in refining the description of intermediate $R$
behavior, including the region of the maximum corresponding to the
\emph{most hydrophobic} solutes.

\begin{figure} 
\includegraphics[scale=0.5]{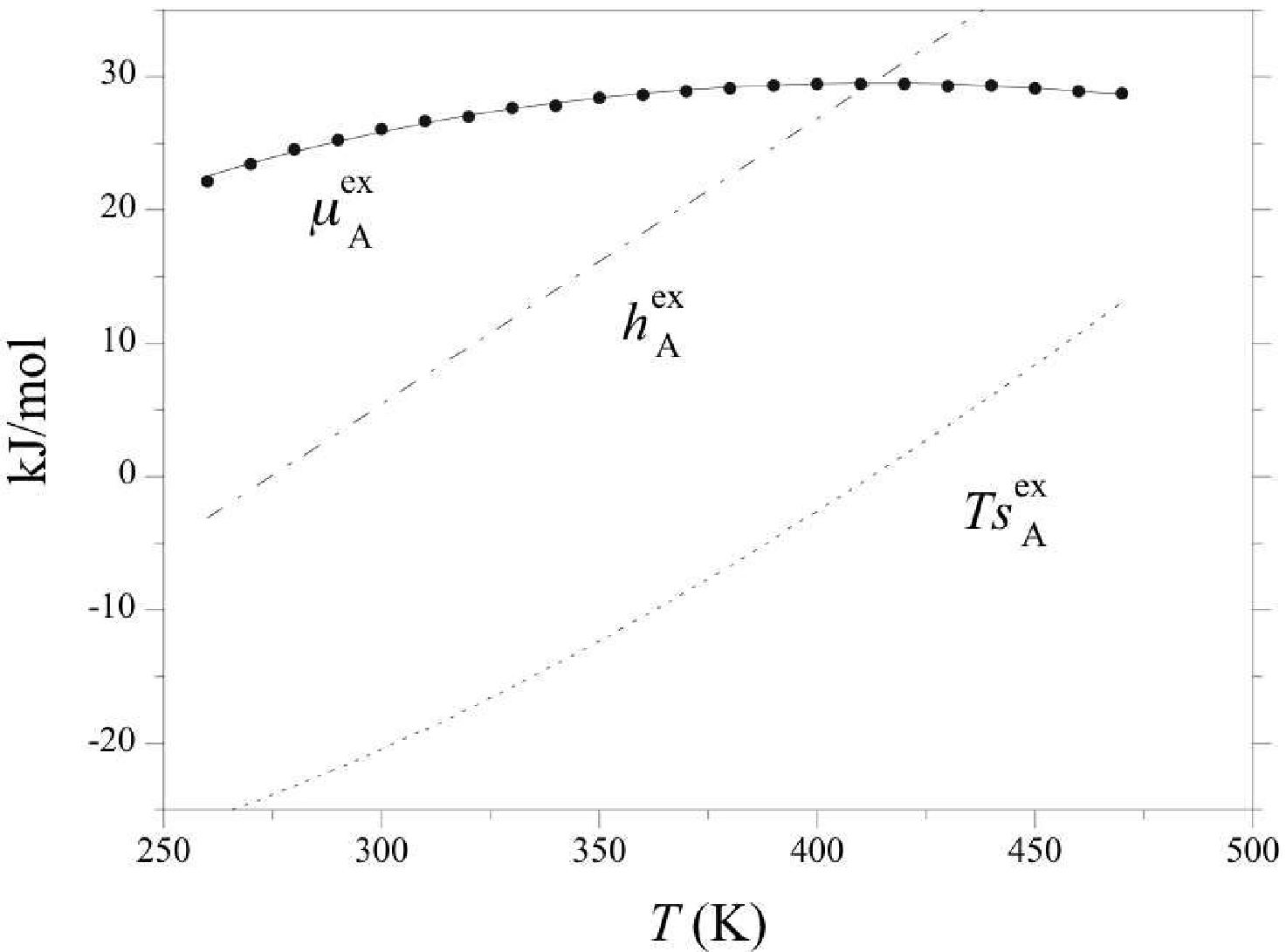}
\caption{Excess chemical potential, enthalpy, and entropy of a methane
sized hard sphere solute (R = 3.3~\AA) in water as a function of
temperature along the saturation curve. The points are the explicit
simulation results for the chemical potential. Error bars are
comparable in size to the points. The curves for the excess chemical
potential, enthalpy, and entropy are labeled in the figure. The curves
were determined under the assumption that the heat capacity is
independent of temperature [Eqs.~\eqref{16}].\label{fig3}}
\end{figure}

\subsection{Hydration thermodynamics of hydrophobic species: temperature
signatures and solubility minima} The hydration free energy of a methane
sized ($R$ = 3.3~\AA) hard-sphere solute  in water as a function of
temperature along the saturation curve is shown in Fig.~\ref{fig3}. The
simulation results for the chemical potential pass through a maximum at
$ T \approx$ 400~K, at which point the hydration entropy defined by
$s_\mathrm{A}^\mathrm{ex}$ =
$-\partial\mu_\mathrm{A}^\mathrm{ex}/\partial T\vert_\mathrm{sat}$ is
zero. To extract the enthalpy and entropy of hydrophobic hydration from
the chemical potential, we assume that the heat capacity $\partial
h_\mathrm{A}^\mathrm{ex}/\partial T\vert_\mathrm{sat}$ = T$\partial
s_\mathrm{A}^\mathrm{ex}/\partial T\vert_\mathrm{sat}$ =
$c_\mathrm{A}^\mathrm{ex}\left( T\right)$ is independent of temperature.
In this case, the hydration enthalpy, entropy, and free energy are
\begin{subequations}
\begin{align}
h_\mathrm{A}^\mathrm{ex} & = h_\mathrm{A}^\mathrm{ex}\left( T_0\right)  +  \left( T-T_0\right)c_\mathrm{A}^\mathrm{ex}\left( T_0\right)~, \\
s_\mathrm{A}^\mathrm{ex} & =  s_\mathrm{A}^\mathrm{ex}\left( T_0\right) 
 +  \ln\left(\frac{T}{T_0}\right)c_\mathrm{A}^\mathrm{ex}\left( T_0\right)~, \\
\mu_\mathrm{A}^\mathrm{ex} & =  
\mu_\mathrm{A}^\mathrm{ex}\left( T_0\right) 
+  \left(T-T_0\right)\left(c_\mathrm{A}^\mathrm{ex}\left( T_0\right) -
s_\mathrm{A}^\mathrm{ex}\left( T_0\right)\right)  \nonumber \\
&\quad\quad -  T\ln\left(\frac{T}{T_0}\right)c_\mathrm{A}^\mathrm{ex}\left( T_0\right)~,
\end{align}
\label{16}
\end{subequations}
respectively, and $\mu_\mathrm{A}^\mathrm{ex}\left( T_0\right)$ =
$h_\mathrm{A}^\mathrm{ex}\left( T_0\right)$ $-$
$T_0s_\mathrm{A}^\mathrm{ex}\left( T_0\right)$. 

The enthalpy and entropy of hydration of the methane-sized hard-sphere
solute solute is shown alongside the fitted free energy in
Fig.~\ref{fig3}. The hydration entropy is negative and unfavorable at
room temperature \cite{Kauzmann:59,Tanford:80,Blokzijl:AC:93}. With
increasing temperature the entropy changes sign indicative of a positive
heat capacity increment. The entropy and heat capacity at $T = $ 298~K
for the hard-sphere solute are $-69.5$ J/(mol K) and 214 J/(mol K), respectively,
which is in excellent agreement with the experimental values for the
entropy and heat capacity of $-66.7$ J/(mol K) and 209 to 237 J/(mol K)
for methane at $T = $ 298~K
\cite{RETTICHTR:SOLOGI,NAGHIBIH:HEAOSO,BENNAIMA:SOLTON,Lazaridis:92}.
Over most of the temperature range considered, the  hydration enthalpy
is positive and unfavorable for hydration, in disagreement with the
experimental enthalpy for methane of $-11.5$ kJ/mol at $T = $ 298~K,
largely a result of the neglect of attractive interactions with water.
The \emph{iceberg} hypothesis of Frank and Evans \cite{Frank:JCP:45}
suggests that local freezing of water in the vicinity of hydrophobic
species contributes to the experimental negative hydration enthalpy. In
the case of the methane sized hard sphere though, the hydration enthalpy
at $T = $ 298~K is 5.0 kJ/mol, contrary to customary expectations.

\begin{figure} 
\includegraphics[scale=0.5]{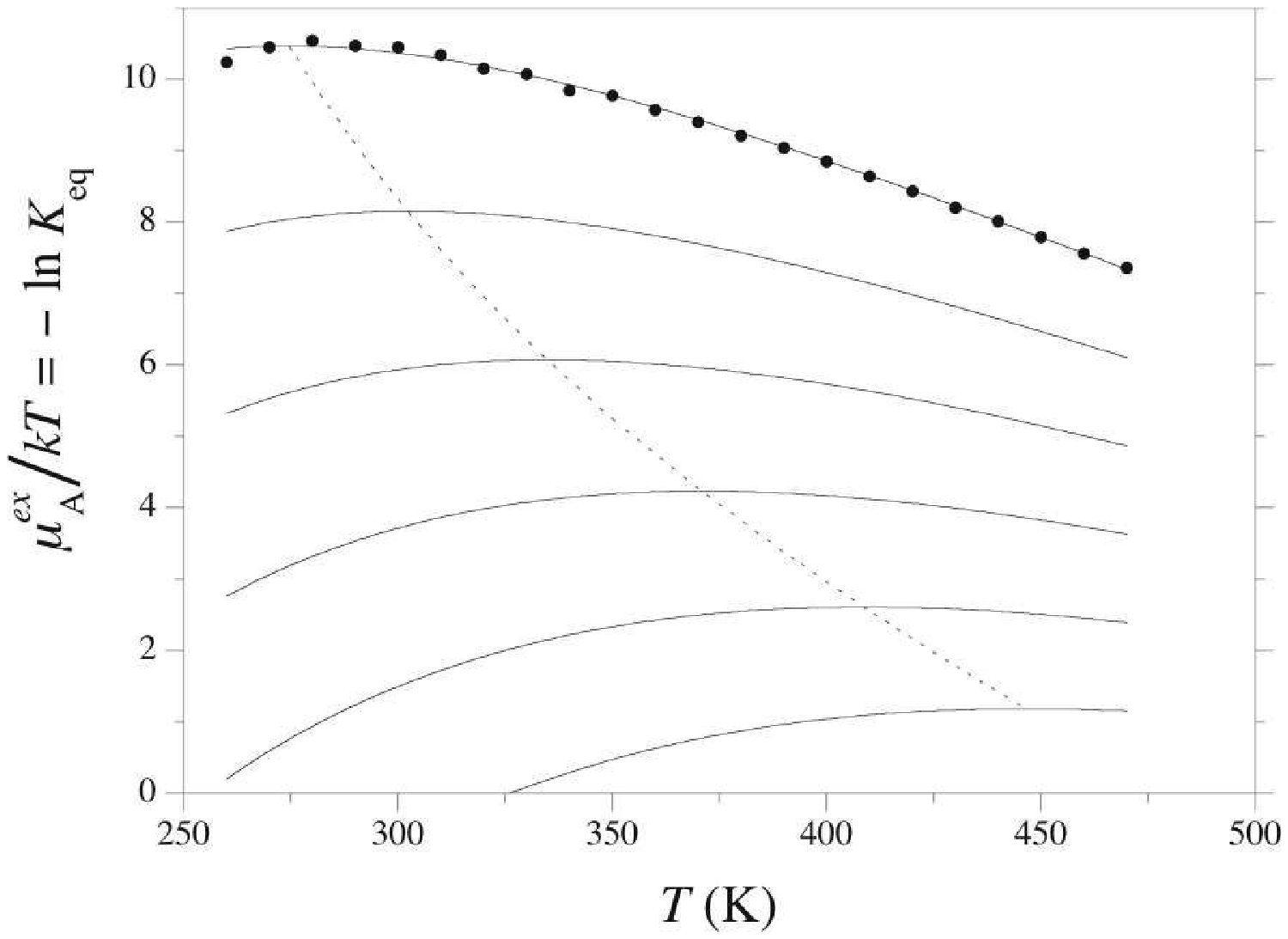}
\caption{Solubility of a 3.3~\AA\ {\em hard-core + weak attractions}
solute as a function of temperature with increasing strength of
attractive interactions modeled as in the van der Waals
equation-of-state, $\mu_\mathrm{A}^\mathrm{ex}$ =
$\left\lbrack\mu_\mathrm{A}^\mathrm{ex}\right\rbrack_\mathrm{HS} -
a_{SW}\rho_\mathrm{W}$. The points are simulation results from particle
insertion probabilities. The solid curves are the solubilities with
lower curves indicating increasing attractive interactions,
$a_\mathrm{SW}\geq$ 0. The dashed curve indicates the maxima in
$\mu_\mathrm{A}^\mathrm{ex}/kT$ with increasing interactions,
corresponding to minima in the Ostwald solubility, defined as
$K_\mathrm{eq} = \exp\left\lbrack - \mu_\mathrm{A}^\mathrm{ex}
\left(\mathrm{aq}\right) /kT\right\rbrack$ [Eq.~\eqref{Ostwald}]. }
\label{fig4}
\end{figure}

The ratio of the chemical potential and $kT$  dictates the Ostwald
solubility. $-\ln K_\mathrm{eq}$ for the methane sized solute in water
as a function of temperature is shown in Fig.~\ref{fig4}. This quantity
passes through a maximum near $T = $ 280~K, corresponding to a minimum
in the solubility. This observation is in agreement with information
theory \cite{Garde:99} and equation-of-state \cite{Ashbaugh:02} models
of hard-sphere solubilities that link the solubility minimum to the
density maximum at $T = $ 277~K for pure water. The solubility minimum
corresponds to the point at which the enthalpy,
$h_\mathrm{A}^\mathrm{ex}$ = $-T^2
\partial\left(\mu_\mathrm{A}^\mathrm{ex}/T\right)/\partial
T\vert_\mathrm{sat}$ equals zero. Real nonpolar solutes display
solubility minima at temperatures well above the density maximum,
largely as a result of attractive interactions between the solute and
water. These interactions,  not included in the present simulations, can
be included approximately by assuming they are proportional to the
density of liquid water, as in the van der Waals equation-of-state. The
resulting chemical potential is $\mu_\mathrm{A}^\mathrm{ex}$ =
$\left\lbrack\mu_\mathrm{A}^\mathrm{ex}\right\rbrack_\mathrm{HS}$ $-$
$a_\mathrm{AW}\rho_\mathrm{W}$ \cite{Garde:99}. The effect of including
solute-water interactions on the solubility is shown in Fig.~\ref{fig4}.
Increasing these interactions systematically shifts the maximum in
$\mu_\mathrm{A}^\mathrm{ex}/kT$, out to greater temperatures, in
agreement with the experimental observation of solubility minima at
higher temperatures.

For the hard-sphere solutes  the simulation results have slightly more
curvature at temperatures near the solubility minimum than predicted by
Eq.~\eqref{16} [Fig.~\ref{fig4}]. While the fit is accurate, the
enhanced curvature suggests the heat capacity is not constant as assumed
above, but is slightly larger at low temperatures. Indeed this has been
observed experimentally \cite{GILLSJ:ANOHOH},  and is borne out by
theoretical models of hydrophobic hydration as well
\cite{Silverstein:2001,Hummer:00,Ashbaugh:02}.  Nevertheless, the
temperature dependence of the heat capacity is  minor, and including it
is a complication of secondary  importance to the interpretations here.
We therefore neglect it in our thermodynamic analysis.

\begin{figure} 
\includegraphics[scale=0.5]{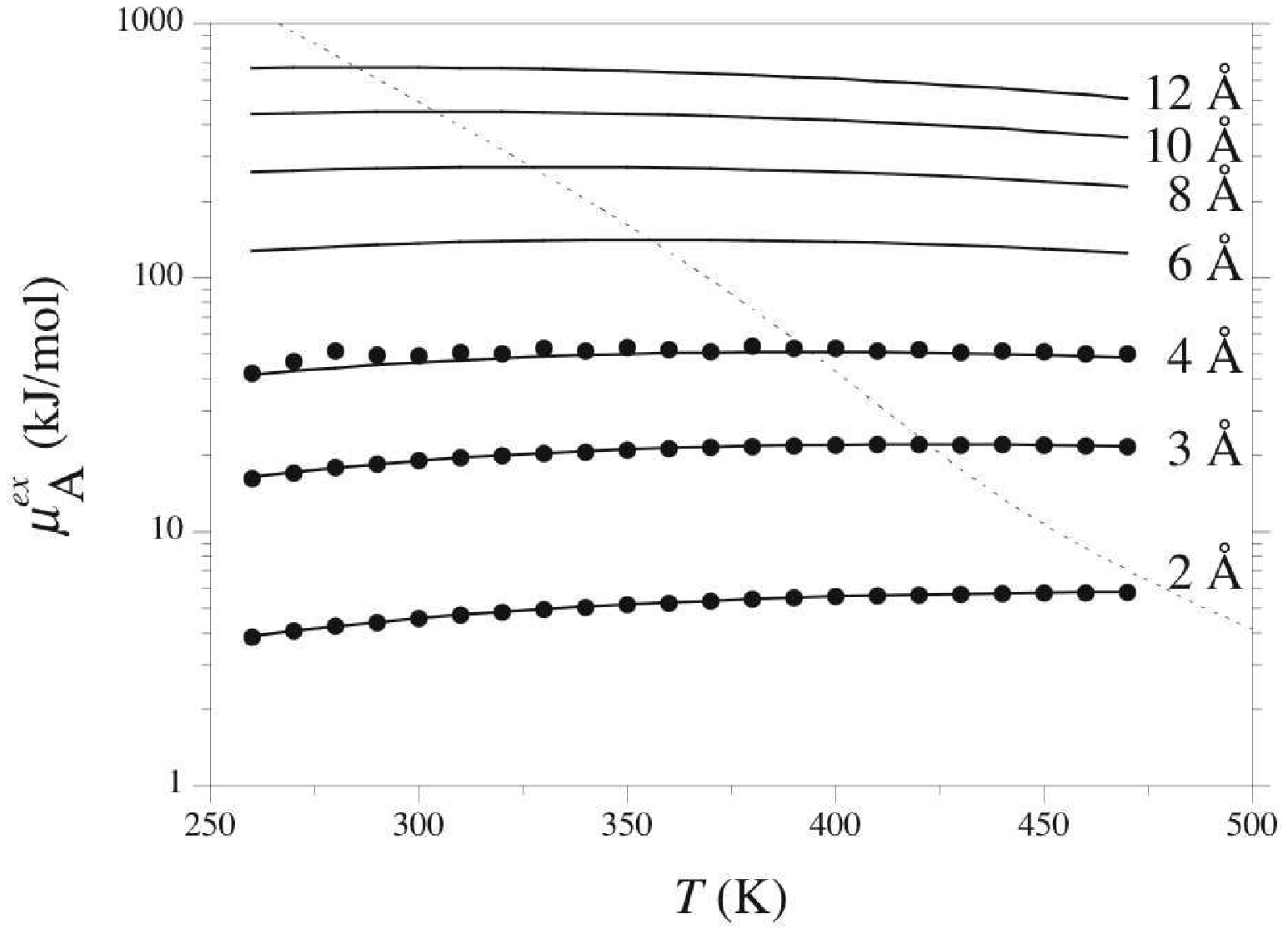}
\caption{Excess solute chemical potential as a function of temperature for
solutes of varying size. The solid lines correspond to the revised SPT
model. The solid circles, open circles, and crosses correspond to
explicit molecular simulation results simulation results for the 2~\AA,
3~\AA, and 4~\AA\ radius solutes, respectively.  Estimated statistical
errors are smaller than the plotting symbols. The
dashed line indicates the locus of chemical potential maxima, where
$s_\mathrm{A}^\mathrm{ex}$ = 0, with changing cavity size.\label{fig5}}
\end{figure} 

Fig.~\ref{fig5} shows how the chemical potential of hard spheres in water
as a function of temperature with increasing solute size. In the range
of sizes shown, the maximum in the chemical potential shifts from
temperatures above $T = $ 470~K, above the window of temperatures simulated, for
the 2~\AA\ radius solute to lower temperatures with increasing solute
sizes. For cavities not much larger than 12~\AA\ (not shown in figure),
the maximum falls below $T = $ 260~K, below the simulation window and the normal
freezing point of water. Thus, for molecularly sized cavities,
hydrophobic hydration is opposed by a negative entropy, at temperatures
below the chemical potential maximum, over most of the range of
temperatures simulated. For meso- and macroscopic cavities, however,
this trend is reversed and hydration is favored by a positive
dissolution entropy but is insoluble as a result of a dominating positive
enthalpy (demonstrated in the following section), dictated
by the temperature dependence of the surface tension of water.

\subsection{Surface area dependence of hydration thermodynamic
properties: curvature corrections and a Tolman length} In an effort to
compare and correlate hydration free energies of a variety species, it
is common to calculate the free energy cost per unit area for hydrating
nonpolar solute surfaces, also referred to as a molecular surface
tension \cite{Tanford:79,HERMANNRB:USEOSC}. This molecular surface
tension, however, is generally not equal to the free energy of creating
a macroscopic flat interface, in part due to curvature and
structural differences between water at molecular and macroscopic
interfaces. Nevertheless, SPT systematically interpolates the surface
tension between these two length-scale extremes, and provides insight
into their relationship \cite{AshbaughHS:Effssa,Huang:00,HuangDM:01}.

Under the assumption that the pressure contribution to the hydration
free energy is negligible, an excellent assumption for liquid water, a
surface tension for hydration of a hard-sphere solute is obtained from
the surface area derivative of the chemical potential
\cite{AshbaughHS:Effssa}. This derivative depends, however, on the
definition of the  surface area. A natural choice for the solute area is
defined by $R$, and is referred to as the solvent accessible surface
(SAS) area, $A_\mathrm{SAS} = 4\pi R^2$. Differentiating the chemical
potential with respect to this surface area yields
\begin{eqnarray}
\gamma_\mathrm{SAS}(R) =
\frac{\partial\mu_\mathrm{A}^\mathrm{ex}}{\partial 4\pi R^2} =\frac{1}{2}kT\rho_\mathrm{W} G(R)R~.
\label{17a}
\end{eqnarray}
More generally, the solute surface can be defined by a radius $R -\Delta
R$ . In this case the surface tension is
\begin{eqnarray}
\gamma\left(R;\Delta R\right) & = &
\frac{\partial\mu_\mathrm{A}^\mathrm{ex}}{\partial \left\lbrack4\pi \left(R-\Delta R\right)^2\right\rbrack} 
=\frac{kT\rho_\mathrm{W} G(R)R^2}{2\left(R-\Delta R\right)} 
 \nonumber \\
 & = & \frac{\gamma_\mathrm{SAS}(R)}{1-\Delta R/R}~.
\label{17b}
\end{eqnarray}

\begin{figure} 
\includegraphics[scale=0.5]{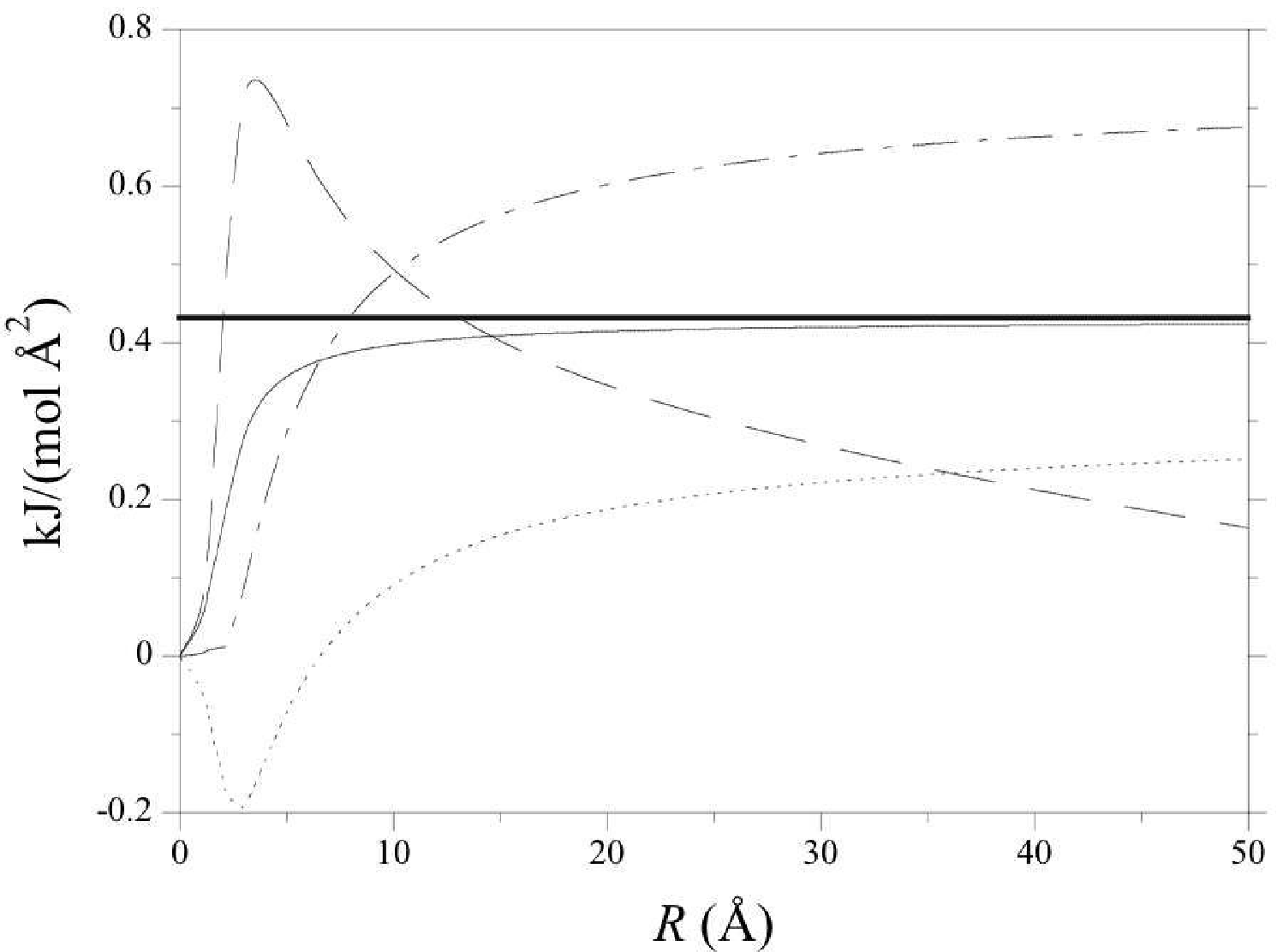}
\caption{SAS area derivatives of the hard-sphere solute hydration
thermodynamics as a function of solute radius at $T$ = 300~K. The thin
solid, long-short dashed, and short dashed, and long dashed lines
correspond to $\partial\mu_\mathrm{A}^\mathrm{ex}/\partial
A_\mathrm{SAS}$, $\partial h_\mathrm{A}^\mathrm{ex}/\partial
A_\mathrm{SAS}$, $T\partial s_\mathrm{A}^\mathrm{ex}/\partial
A_\mathrm{SAS}$, $T\partial c_\mathrm{A}^\mathrm{ex}/\partial
A_\mathrm{SAS}$, respectively. The thick horizontal line indicates the
macroscopic surface tension for a flat surface. \label{fig6}}
\end{figure}

The SAS  tension as a function of $R$ at $T = $ 300~K is shown in
Fig.~\ref{fig6}. The surface enthalpy, $\partial
h_\mathrm{A}^\mathrm{ex}\left( T_0\right)/\partial A_\mathrm{SAS}$,
entropy, $\partial s_\mathrm{A}^\mathrm{ex}\left( T_0\right)/\partial
A_\mathrm{SAS}$, and heat capacity, $\partial
c_\mathrm{A}^\mathrm{ex}\left( T_0\right)/\partial A_\mathrm{SAS}$, are
included in this figure. For small cavities, all the surface
thermodynamic properties go to zero as $R\rightarrow$ 0. With increasing
size, the surface tension increases monotonically and approaches its
asymptotic limit for a flat interface of $\gamma_\infty$ = 0.432 kJ/(mol
\AA$^2$) = 71.7dyne/cm. The other surface properties, most notably the
heat capacity, approach their asymptotic plateaus more slowly with
increasing $R$. Like the surface tension, the surface enthalpy
monotonically increases with increasing solute size. The surface entropy
and heat capacity, on the other hand, vary in distinctly different ways
for molecular and macroscopic surfaces, indicating changes in the
mechanism of hydration \cite{Southall:00}. In particular, the surface
entropy is initially negative beginning from $R$ = 0, consistent with
the experimental thermodynamics of hydrophobic hydration for molecular
solutes, reaches a minimum at $R \approx $ 3.3~\AA\ and then increases,
eventually becoming positive as expected from the temperature dependence
of the liquid-vapor interface. It is curious that the  size at this
minimum is close to the size of the maximum of $G(R)$, that is for the
\emph{most hydrophobic} hard sphere solute. While the surface heat
capacity is positive over the entire size range, it reaches a maximum at
solute radii comparable to the position of the minimum in the entropy,
suggesting the two are related.  Moreover, we may infer that the maxima
in the surface entropy and heat capacity are linked to the breakdown of
the aqueous network in the vicinity of the hard-sphere solute as
observed in simulation studies linking solute-water correlations to the
thermodynamics of hydrophobic hydration \cite{LAZARIDIST:SIMSOT}.

\begin{figure} 
\includegraphics[scale=0.3]{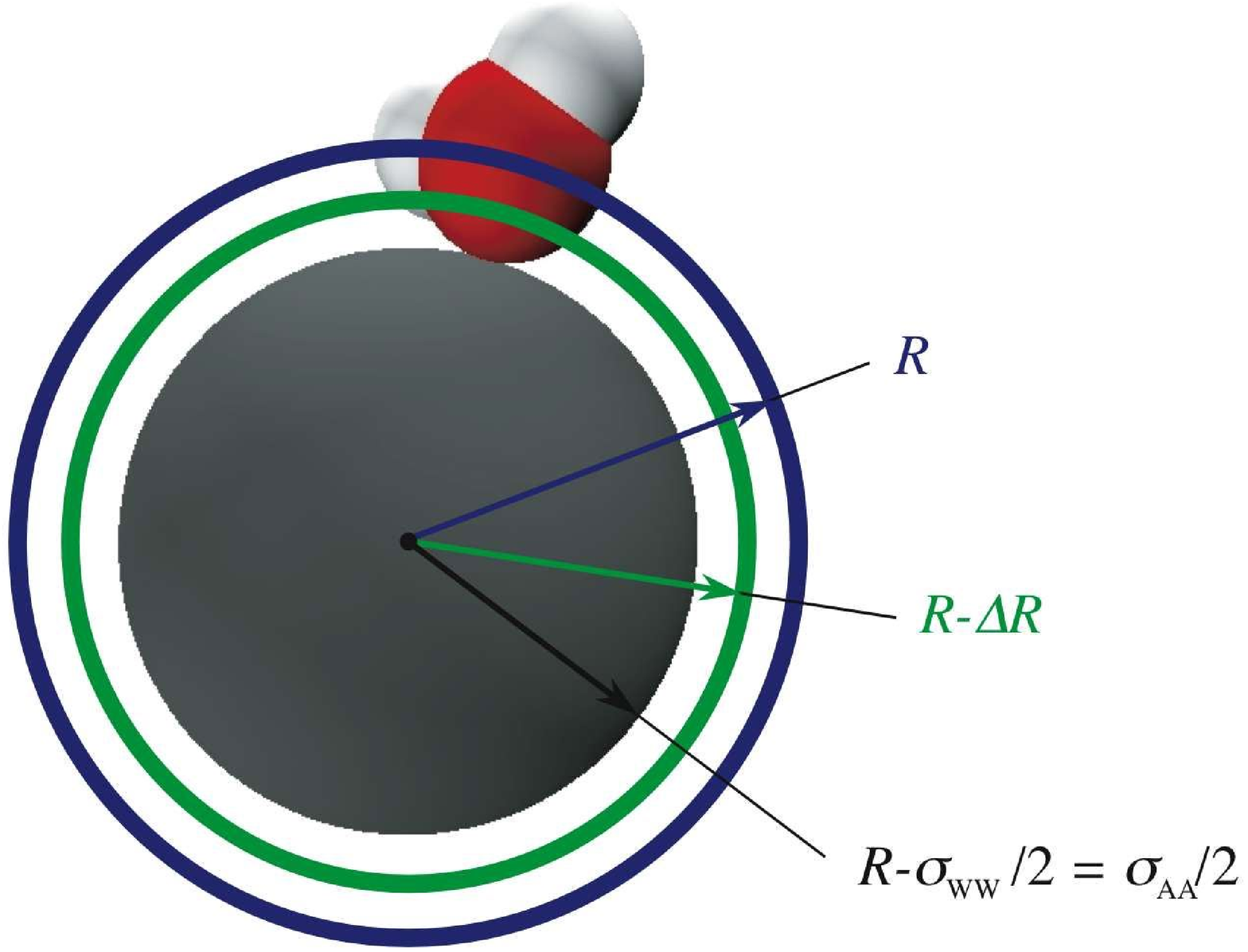}
\caption{Alternative definitions of the radius of a hard sphere cavity.
The solvent accessible radius, $R$, is given by the distance of closest
approach between the center of the cavity and the water oxygen center. 
The radius $R - \sigma_\mathrm{WW}/2$ demarks the van der Waals boundary
of the cavity at $\sigma_\mathrm{AA}/2$.  $R - \Delta R$ locates a
neighboring surface that might provide a  curvature corrected surface
tension. \label{fig12}}
\end{figure}

Sharp and coworkers \cite{Sharp:91} suggested that rather than relying
solely on the SAS to determine the molecular surface tension, this
tension needs to be corrected for the curvature of the molecular
interface to reconcile the difference between molecular and macroscopic
surface tension.  Based on geometric arguments, they proposed that the
radius of a water molecule is the length-scale over which this
correction must be applied. In effect, their work indicates that the van
der Waals surface, i.e., $\Delta R$ = $\sigma_\mathrm{WW}/2$ = 1.4~\AA,
provides a superior description of molecular solute hydration
\cite{Sharp:91,Jackson:94}.  A schematic illustration of the van der
Waals, solvent accessible, and curvature corrected radii of a hard
sphere solute in water is given in Fig. ~\ref{fig12}.  For a methane
sized solute  $\gamma\left(R=3.3\mathrm{\AA};\Delta R = 0.0
\mathrm{\AA}\right)$ = 0.300 kJ/(mol \AA$^2$), 30\% lower than the
macroscopic value, while $\gamma\left(R=3.3\mathrm{\AA};\Delta R =
1.4\mathrm{\AA}\right)$ = 0.521 kJ/(mol \AA$^2$), 20\% greater than the
macroscopic value. While neither of these two surfaces gives the
macroscopic result, they do bracket $\gamma_\infty$ suggesting there an
optimal intermediate value of $\Delta R$ for which the surface tension
is size independent. 

\begin{figure} 
\includegraphics[scale=0.5]{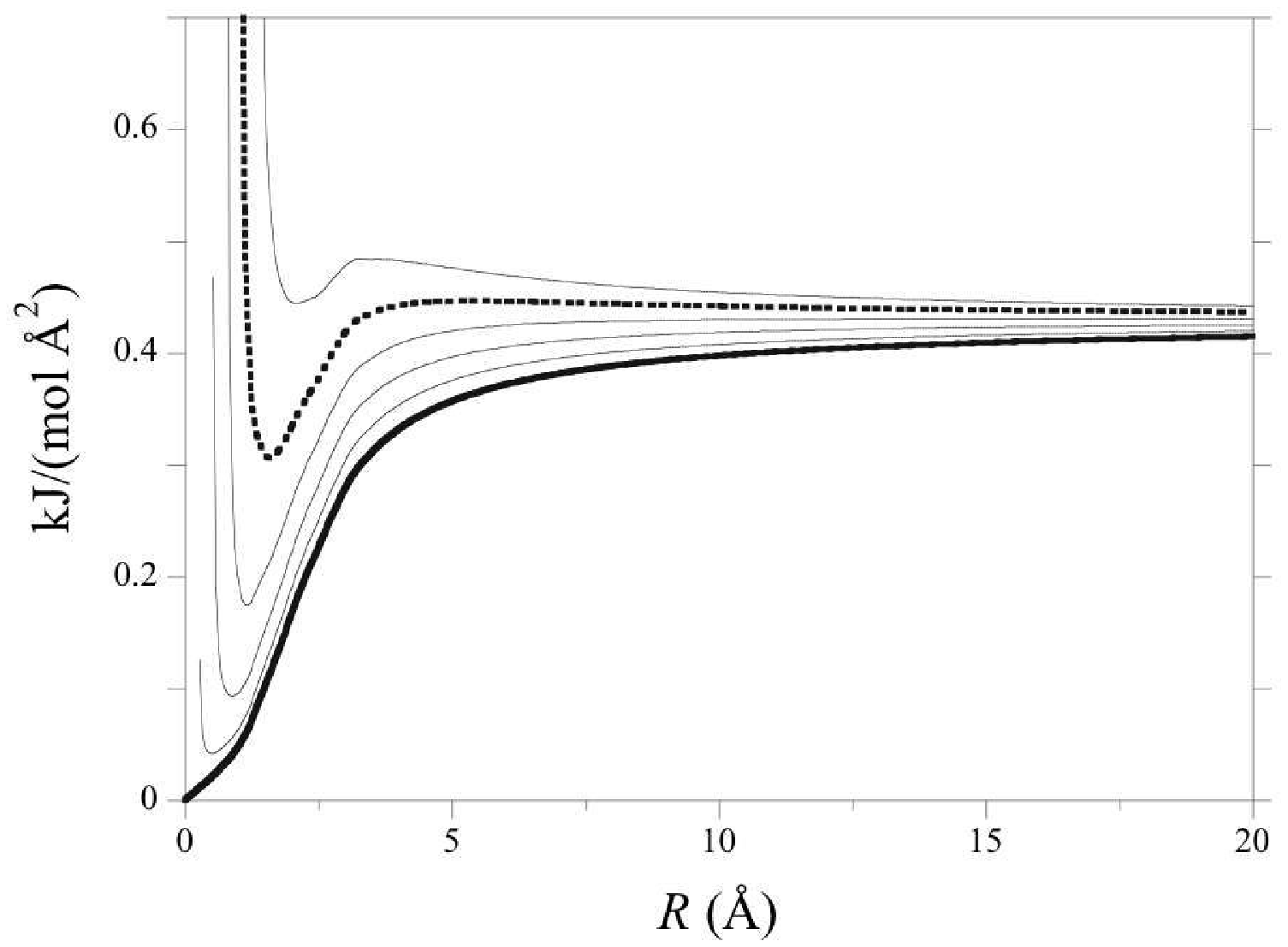}
\caption{Surface tension for hydration of hard-sphere solutes as a function of
solute radius at $T = $ 300~K employing different definitions of the solute
surface. The thick solid line corresponds to the surface tension
determined by the derivative with respect to the solvent accessible
surface area defined by the radius $R$ [Eq.~\eqref{17a}]. The lines above
the baseline SAS surface tension indicate the effect of increasing
$\Delta R$ in 0.25~\AA\ increments from 0.25~\AA\ to 1.25~\AA\
[Eq.~\eqref{17b}]. The thick dashed line corresponds to $\Delta R$ =
1~\AA.\label{fig7}}
\end{figure}

Fig.~\ref{fig7} shows how  the surface tension varies with  $\Delta
R$. The divergence in $\gamma\left(R;\Delta R\right)$ results from the
divergence in Eq.~\eqref{17b} as $R\rightarrow\Delta R$. For solutes
larger than 2.5~\AA, using $\Delta R$ = 1~\AA\ yields a surface tension
that is only weakly size dependent. Indeed,
$\gamma\left(R=3.3\mathrm{\AA};\Delta R = 1.0 \mathrm{\AA}\right)$ =
0.430 kJ/(mol \AA$^2$) is in excellent agreement with the macroscopic
value, suggesting that Honig and coworker's geometric estimate of the
curvature correction length-scale is correct. This argument falls apart,
however, when we consider the temperature dependence of the Tolman
length.

\begin{figure} 
\includegraphics[scale=0.5]{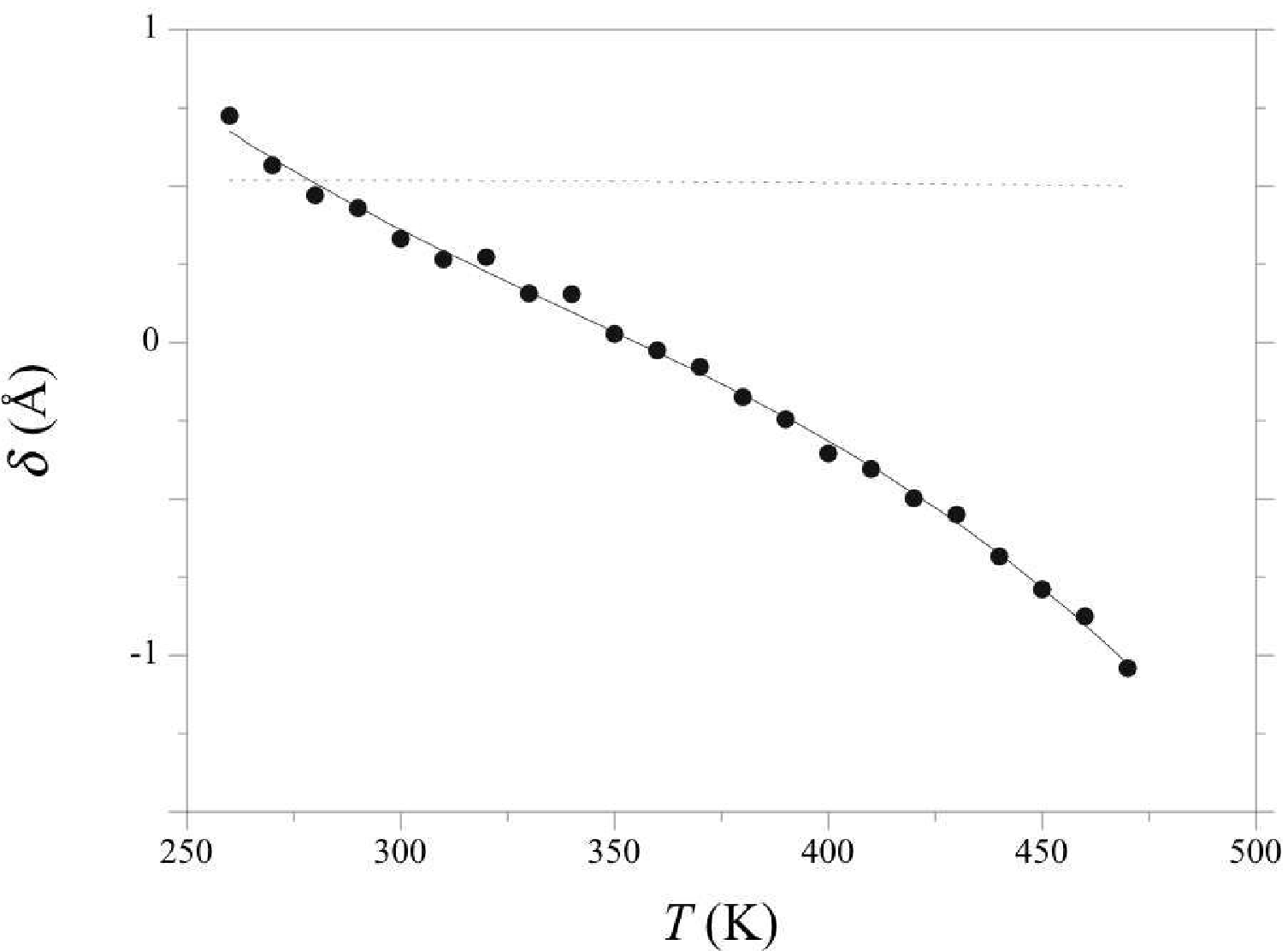}
\caption{The  curvature correction, $\delta$, as a function of temperature along the
saturation curve of water. The points correspond to the values
determined by the fit of Eq.~\eqref{rrspt-mu.0} to the simulation free
energies. The solid line is a guide to the eye for the fitted
results. The dashed line corresponds to the classic SPT prediction for
the Tolman length, Eq.~10 in \cite{Stillinger:73}.
\label{fig8}}
\end{figure}

The Tolman length $\delta$, described in Sec.~\ref{sec:classic}, can be
accessed by the revised SPT through Eqs.~\ref{eq:tolman} and~\ref{17a}:
\begin{eqnarray}
\gamma_\mathrm{SAS}\left( R \right) \sim \gamma_\infty \left(1  - \frac{2\delta}{R}\right)~.
\end{eqnarray} 
Substituting this expression into Eq.~\eqref{17b} yields
\cite{AshbaughHS:Effssa}
\begin{eqnarray}
\gamma\left(R;\Delta R\right)  \sim \gamma_\infty \left(
\frac{1 -2\delta/R}{1-\Delta R/R}
\right)
\end{eqnarray}
for the surface tension referenced to a surface displaced by $\Delta R$.
 Thus, for large $R$ the optimal surface for obtaining a size
independent surface free energy is $\Delta R$ = $2\delta$.   This Tolman
length can be calculated from classic SPT, see Eq.~10 in
\cite{Stillinger:73}, and yields an nearly temperature independent
$\delta  \approx$ 0.5 \AA\ [Fig.~\ref{fig8}], in good agreement with the
empirical $\Delta R$ at $T = $ 300~K obtained above. The revised SPT,
however, finds that $\delta$ is strongly temperature dependent. While
the classic SPT correctly predicts the magnitude of $\delta$ at low
temperature, $\delta$ decreases with temperature and changes sign near
$T = $ 350~K. Furthermore, \cite{MoodyMP:Curdst} suggest that the Tolman
length might change sign also for a Lennard-Jones solvent.  Thus
assuming the $\Delta R$ is simply dictated by the size of a water
molecule leads to fundamentally flawed interpretations of the
relationship between molecular and macroscopic surface tensions
\cite{Sharp:91,Jackson:94}. In retrospect, the temperature dependence of
the curvature correction might have been anticipated by the entropic
differences between hydrating a molecular and mesoscopic interface and
the significantly different temperature dependencies of the associated
surface thermodynamic properties [Fig.~\ref{fig6}]. The curves observed
in this figure are simply too rich to be described by a temperature
independent length-scale.

\subsection{Entropy convergence and solute size} When the hydrophobic
component of the hydration entropies of these molecules is extrapolated
to high temperatures,  the entropies  converge to one another within a
narrow range near $T = $ 385~K
\cite{Privalov:79,Baldwin:86,PRIVALOVPL:STAOPA,Murphy:90}. This
phenomenon of entropy convergence is  feature of hydrophobic hydration
believed to be shared between small molecule hydration and protein
unfolding thermodynamics. Baldwin and Privalov noted that this can
result from proportionality of the entropy and heat capacity of
hydration to one another. The most successful explanations for the
convergence temperature for small molecules have related the convergence
temperature to the equation-of-state of pure water
\cite{Garde:96,Hummer:98,Garde:01,Ashbaugh:02}.  \cite{Huang2:00} argued
that for species larger than $R \approx$ 10~\AA\ entropy convergence
does not occur, and therefore proteins do not exhibit this phenomenon .

\begin{figure} 
\includegraphics[scale=0.5]{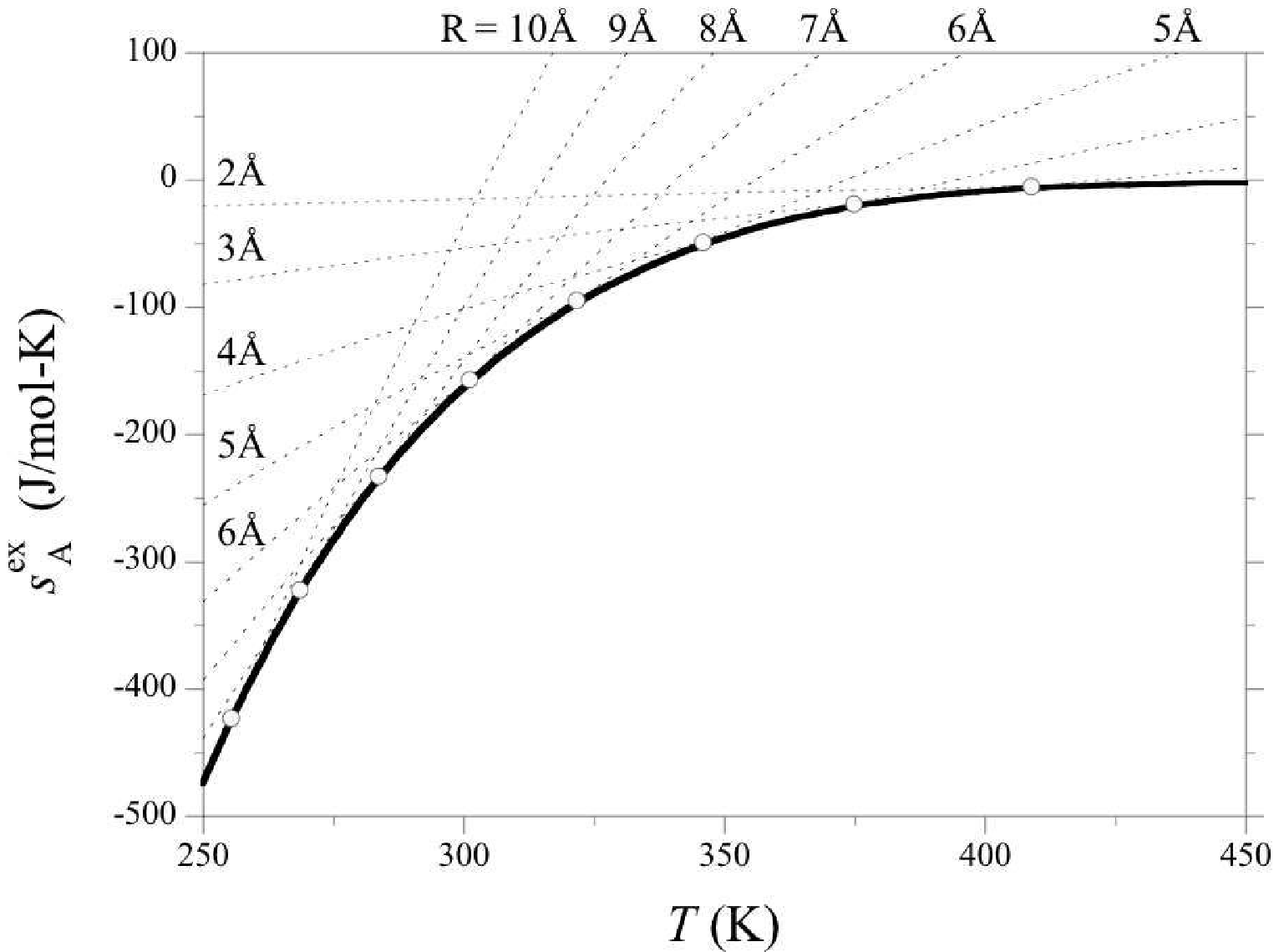}
\caption{Entropy of hydrophobic hydration as a function of temperature
for solutes in the size range 2~\AA$<R<$10~\AA\ in 1~\AA\ increments. The
dashed lines are the excess entropies of hydration. The open circles are
the convergence temperatures for consecutive solutes, \emph{i.e.},
$s^\mathrm{ex}_\mathrm{A}(R)$ = $s^\ast_\mathrm{A}(R + 1\mathrm{\AA})$.
The thick solid line indicates the entropy convergence temperature in
the limit of infinitesimal perturbations in $R$.\label{fig9}}
\end{figure}

In Fig.~\ref{fig9} we have plotted the hydration entropies for hard-sphere
solutes as a function of temperature for solutes  in the size range
2~\AA\ $\leq  R \leq $ 10~\AA. Indeed, there is a broad range of
convergence temperatures observed at each point where an entropy curve,
for a given size, crosses that for another solute. For example, the
2~\AA\ solute entropy intersects the 3~\AA\ solute entropy at $T = $
410~K, while the 2~\AA\ solute entropy intersects the 10~\AA\ solute curve
at $T = $ 300~K, indicating that there is no unique convergence
temperature. Experimental identifications of that entropy convergence,
however, have largely concerned themselves with solutes similar in size.
We therefore consider how the convergence temperature changes with
differential perturbations in the solute size. In this case, convergence
occurs at the temperature for which $\Delta
s_\mathrm{A}^\mathrm{ex}\left(R\rightarrow R + \delta R\right)$ = 0.
Assuming the hydration heat capacity is independent of temperature,
$T_\mathrm{c}$ is determined by the relationship
\begin{eqnarray}
T_\mathrm{c} = T_0 \exp\left( - \frac{\partial s_\mathrm{A}^\mathrm{ex}\left( T_0\right)}{\partial
A_\mathrm{SAS}}/\frac{\partial c_\mathrm{A}^\mathrm{ex}\left( T_0\right)}{\partial A_\mathrm{SAS}}\right)
\label{econv}
\end{eqnarray}
with a size dependence dictated by the relationship between $\partial
s_\mathrm{A}^\mathrm{ex}\left( T_0\right)/\partial A_\mathrm{SAS}$ and $\partial c_\mathrm{A}^\mathrm{ex}\left( T_0\right)/\partial A_\mathrm{SAS}$ on
the solute radius in Fig.~\ref{fig6}. The differential entropy
convergence temperature curve determined by these expressions is shown
in Fig.~\ref{fig9}. Two points of interest are immediately apparent:
first the convergence entropy is negative and becomes more negative with
increasing solute size, second the convergence entropy curve more or
less forms a lower bound on the hydration entropy as a function of
temperature, although this is approximate.
\begin{figure} 
\includegraphics[scale=0.5]{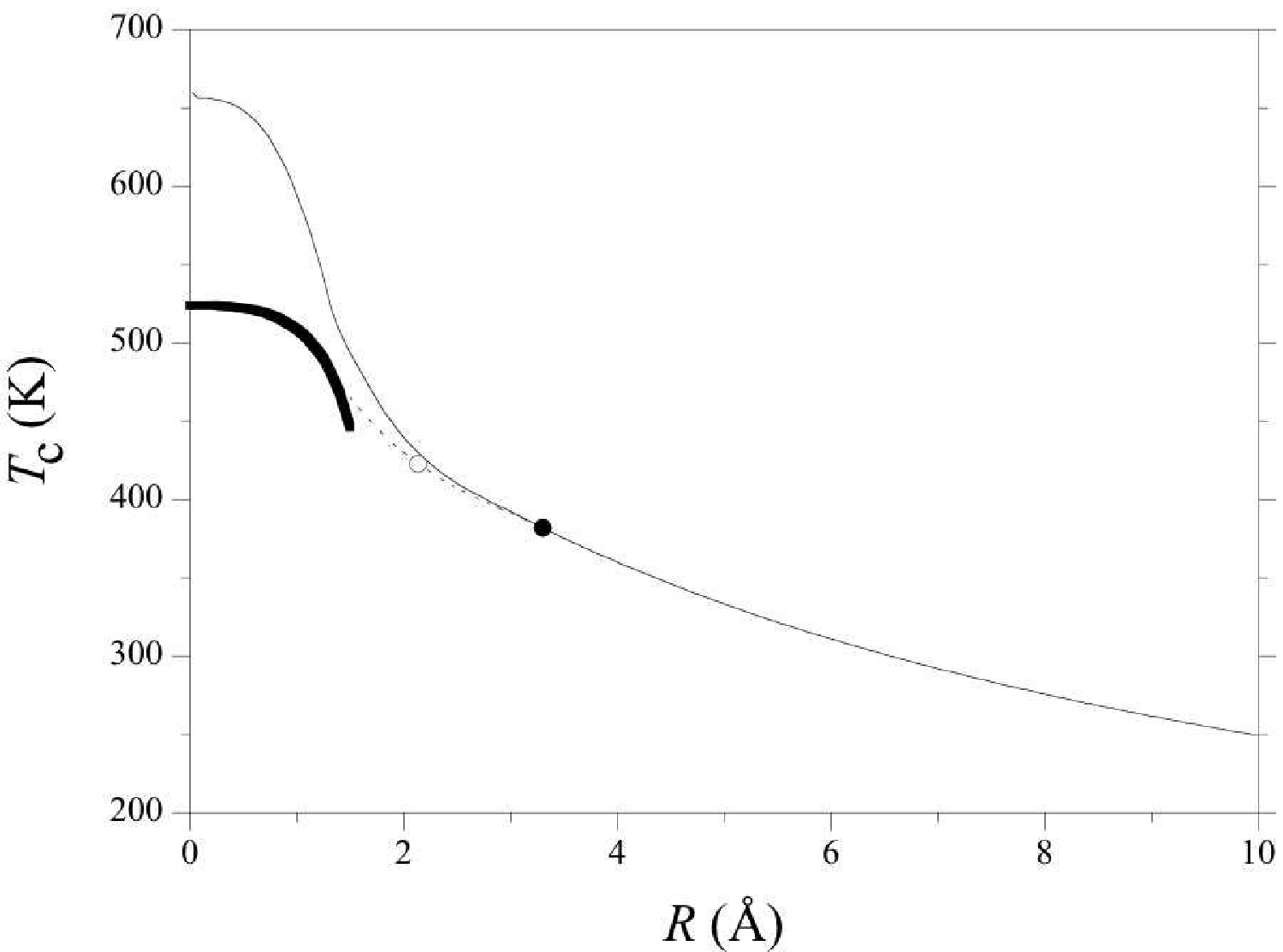}
\caption{Variation of the entropy convergence temperature with
increasing hard-sphere radius. The thin solid line is the convergence
temperature determined under the assumption the heat capacity is
independent of temperature. The thick solid line is the exact entropy
convergence temperature for $R<\sigma_\mathrm{WW}/2$ from
Eq.~\eqref{small-Tc}. The dashed line smoothly interpolates between the
exact and constant heat capacity curves at 1.25~\AA\ and 3.3~\AA,
respectively. The filled circle indicates the entropy convergence
temperature of a methane-sized solute ($T_\mathrm{c} = $ 382~K). The
open circle indicates the entropy convergence temperature based on the
IT criterion ($T_\mathrm{c} = $ 420~K).\label{fig10}}
\end{figure}

One of the implications of Eq.~\eqref{econv} is that if the entropy is a
linear function of the heat capacity, \emph{i.e.}, $s^\mathrm{ex}\left(
T_0\right)$ = $m c^\mathrm{ex}\left( T_0\right) + b$, as suggested by
\cite{Baldwin:86,Murphy:90},  then the convergence temperature would be
independent of solute size. In Fig.~\ref{fig10} we see that
$T_\mathrm{c}$ has a significant solute size dependence, indicating that
this assumption has limited validity. For solutes approaching zero
radius, $T_\mathrm{c}$ plateaus at a maximum. With increasing solute
size $T_\mathrm{c}$ decreases so that above $R \approx $ 8~\AA\ it is less
than the normal freezing point of water. At the intermediate methane
radius of 3.3~\AA, however, the convergence temperature is 382K in
excellent agreement with the experimental convergence temperature of
$T = $ 385~K for simple nonpolar gases and linear alkanes.

We note that the convergence temperature as $R\rightarrow$ 0 plateaus at
$T=$ 655~K, above the critical temperature of water at 647~K. This unphysical
result is due, in part, to our extrapolation of the entropies beyond the
range, 260~K to 470~K, to which we fitted Eq.~\eqref{16}. For cavities 
small enough that only one water molecule can fit inside, the free energy is
given by Eq.~\eqref{small-p}, and entropy convergence occurs for
the temperature at which
\begin{eqnarray}
\alpha_\mathrm{sat} T_\mathrm{c} = 1- \frac{4\pi}{3} \rho_\mathrm{W} R^3 ~,
\label{small-Tc}
\end{eqnarray}
where $\alpha_\mathrm{sat}$ = $-\left(\partial \ln
\rho_\mathrm{W}/\partial T\right)_\mathrm{sat}$ is the thermal expansion
coefficient of liquid water along the saturation curve. When this
criterion is applied, it displays a plateau at a physically more
realistic temperature of $T_\mathrm{c}\approx $ 525~K
[Fig.~\ref{fig10}]. With increasing solute size, Eq.~\eqref{small-Tc}
indicates a sudden decrease in $T_\mathrm{c}$ above $R \approx $ 1~\AA,
comparable to that  obtained on the assumption that the heat capacity is
temperature independent. Above radii of 1.25~\AA, Eq.~\eqref{small-Tc}
breaks down as multi-particle correlations began to play a role in the
solute entropy. At this radius, however, the convergence temperatures
are now within the range of temperatures simulated and the application
of Eq.~\eqref{econv} becomes more accurate. It is reasonable then to
interpolate between the convergence temperatures determined by
Eq.~\eqref{econv} and Eq.~\eqref{small-Tc}, as indicated in
Fig.~\ref{fig10}. 

We note that the information theory \cite{Garde:96,Hummer:98}, with
natural simplifying assumptions, indicates entropy convergence when
$T_\mathrm{c} \approx\left(2 \alpha_\mathrm{sat}\right)^{-1}$   = 420~K.
This corresponds on Fig.~\ref{fig10} to a solute radius of $R \approx $
2.1~\AA\, placing this estimate among  small-solute theories. Relaxing
the assumptions used to arrive at this IT criterion lowers the
convergence temperature prediction for methane sized solutes to
$T\approx $ 390~K, improving agreement with the present result of 382~K.

The apparent convergence of the entropy change at $T = $ 385~K to a
value close to zero for a range of hydrophobic solutes and proteins has
suggested that the hydration of these chemically distinct
solutes are related \cite{Baldwin:86,Murphy:90}.  Assuming that the
temperature dependence for protein unfolding arises solely from the
exposure of hydrophobic side chains to water, phenomenological models
have been developed which separate out residual temperature independent
components of the entropy, from contributions such as changes in the
chain conformation, by extrapolating to the convergence temperature.
This procedure relies, in part, on the assumption that the hydration of
surface hydrophobic groups is the same in the native and denatured
conformations, and therefore cancels in the entropy difference.

The surprising  explanation of the entropy convergence phenomena
that tied it to the particular equation of state of liquid water
\cite{Garde:96,Hummer:98,Garde:01,Ashbaugh:02} resolved an important
conundrum for our molecular understanding of hydrophobic effects.
Whether and how this entropy convergence phenomena is involved  with
protein {\em folding} is yet an outstanding question.  Proteins   are
complicated molecules participating in both hydrophobic and hydrophilic
interactions with the solution.  The widely appreciated point that
protein folding thermodynamics may be primarily sensitive to hydration
of unfolded configurations is just as important
\cite{Paulaitis:02,PrattLR:Hydeam}. Considering unfolded  possibilities,
the sizes of the obvious hydrophobic units are in the range of small
molecule hydrocarbon solutes. The largest hydrophobic side chain
---phenyl alanine --- is an example. \cite{Pratt:02} emphasized
 that solution thermodynamic data  are available for
hydrophobic solutes of just this size, {\em e.g.} for benzene,  toluene,
and ethyl benzene \cite{Privalov:89}, and those data suggest that these
solutes exhibit conventional entropy convergence behavior. Thus, it  is
a plausible hypothesis that entropy convergence will be expressed in
protein folding thermodynamics primarily through contributions
associated with the unfolded configurations.

\cite{Huang2:00}  suggested that the hydration of surface nonpolar
groups is better described by the hydration entropy of a solute surface
comparable in size to the protein radius on the order of tens of
angstroms, rather than treating the surface groups individually as
having sizes comparable to simple hydrophobic units. In this hypothesis
it is presumed that entropic contributions for hydrating large
hydrophobic surfaces with attractive dispersion interactions and vicinal
polar/charged groups is the same as that for a hard repulsive surface.
Recent simulations of convex methane clusters have found that when
realistic attractive interactions between water and methane are
included, water packs around the cluster methane sites just as it does
around a solitary methane in solution \cite{AshbaughHS:Effssa,
ChauPL:Comshh}. Moreover, Cheng and Rossky \cite{Cheng:98} found that
the orientational correlations between water and proximal hydrophobic
residues on the convex surfaces of the bee venom protein, meletin, are
comparable to those near individual solitary hydrophobic groups in
solution.  These observations suggest that the available configurational
space, and by extension the entropy, for water molecules near realistic
surface hydrophobic units is the same in the folded and unfolded states,
supporting the assumptions of the phenomenological folding models.  We
note, however, that in the same study Cheng and Rossky found that water
molecules proximal to hydrophobic residues in flat portions of meletin
were more orientationally disordered as a result of the difficulties
associated with maintaining the aqueous hydrogen-bonding network near
restrictive solute topologies \cite{Cheng:98}.  Thus, the applicability
of the phenomenological unfolding model may be complicated by the
protein surface topography and the impact of hydrophobic pockets on the
overall unfolding entropy.  This can introduce further scatter into
measured folding entropies  \cite{Robertson:97}.

\section{Summary and conclusions}

The revised scaled-particle theory bridges the known molecular and
macroscopic limits by utilizing simulation information on multi-body
correlations in liquid water together with  experimental thermodynamic
properties of pure water to construct a functional form for the
hydration free energy of hydrophobic hard-sphere solutes in water. The
classic scaled-particle theory \cite{SWMayer63,BenNaim:67,Stillinger:73}
incorrectly predicts an increase in the surface tension of water with
increasing temperature, and a suspiciously temperature independent
Tolman length which does not agree with revised scaled-particle theory
observations. As a result, application of classic scaled-particle theory
to hydration free energies is largely a fitting exercise to obtain the
effective van der Waals diameter, $\sigma_\mathrm{WW}$, of water.
Conclusions drawn from this tack have weak significance regarding the
molecular origins of the hydrophobic effect, and are limited to
comparisons of the size parameter for water relative to other solvents,
neglecting further molecular detail or specific temperature signatures
of hydrophobic hydration.

The revised scaled-particle theory is more successful, but the success
of the scaled particle approach  derives generally from the remarkable
fact that the results identify a molecular length, near 3.0~\AA, that
provides a good joining point for microscopic and macroscopic
descriptions. The corresponding results for comparative organic solvents
are less simple  \cite{Pratt:92,GrazianoG:Oncsd}.  That micro-macro
joining radius exhibits interesting temperature variation; an accurate
description of those temperature variations is an important part of the
higher fidelity of the revised scaled-particle results.  The revised
scaled-particle theory reproduces the solubility minimum behavior for
small hydrophobic solutes, and demonstrates significant changes in the
hydration mechanism of hard-sphere solutes with increasing solute size.
Specifically,  the hydration thermodynamics of small solutes is
predominantly entropic at room temperature, but the hydration of
mesoscopic cavities is entropically favorable and opposed by a
dominating hydration enthalpy. While it is tempting to describe these
changes in hydration thermodynamics in terms of aqueous hydrogen-bonding
near the hydrophobic entity --- and that can be plausible in the
appropriate theoretical setting --- the scaled-particle theory provides
little in the way of information on the integrity of hydrogen bonded
networks. 

Nevertheless, the revised scaled-particle theory does provide
thermodynamic information that challenges phenomenological views of
hydrophobic effects, particularly the cherished \emph{iceberg}
hypothesis. Whereas the iceberg hypothesis suggests that local freezing
of water molecules in the vicinity of hydrophobic solutes is a source
for the negative hydration enthalpies, we find that at room temperature
the hydration of solutes comparable in size to simple nonpolar gases is
actually unfavorable from an enthalpic as well as an entropic
standpoint.  The experimentally determined favorable enthalpies of
solution of hydrophobic species then are a consequence of attractive
solute-water interactions and not enhanced water-water structuring.

On a molecular level there is a surface that maps macroscopic surface
tensions to molecular values. This reduces the reconciliation of
molecular and macroscopic values of the surface tension to a program of
finding the appropriate dividing surface. The utility of that program
rests on the optimistic expectation that the Tolman length is largely
temperature independent. But that the Tolman length was found to have a
significant temperature dependence in water, changing from positive to
negative at $T \approx $ 350~K, a possibility anticipated by Stillinger
\cite{Stillinger:73}.  As a result then, though the optimal surface for
the description of hydration may be approximated by the solute molecular
surface at low temperatures \cite{Ashbaugh:99,AshbaughHS:Effssa}, with
increasing temperature this optimized surface moves out to the solvent
accessible surface at $T \approx $ 350~K, and ultimately extends beyond
this surface at even higher temperatures as a result of the nontrivial
temperature dependence of the hydration thermodynamics for
molecular-sized solutes.

Finally, the revised scaled-particle theory provides detailed
information on the curious entropy convergence behavior observed for
small molecule solutes and the size dependence of the convergence
temperature. A suitably defined entropy convergence temperature retreats
below the freezing temperature of water for hard spheres the size of
globular soluble proteins.  But heterogeneity of protein-water
interactions and of sizes of hydrophobic units also contribute
importantly to experimental blurring of entropy convergence behavior in
protein unfolding thermodynamic data. Equally important, entropy
convergence behavior for protein {\em folding} thermodynamics may be
primarily expressed through contributions associated with the unfolded
configurations, and due to hydration of hydrophobic side chains of size
corresponding to studied small molecule solutes.

\section*{Acknowledements} This work was supported by the U. S.
Department of Energy, contract E-7405-ENG-36. HSA acknowledges support
from a Los Alamos Director's Fellowship. This research has benefited
from conversations with D. Asthigiri, A.~E. Garc\'{\i}a, S. Garde, G.
Hummer, M.~E. Paulaitis, and A. Pohorille. We  also  thank J. R.
Henderson and R. Evans for their insightful remarks with respect to the
wetting of interfaces.

\bibliographystyle{apsrmp}

\newpage

\end{document}